\documentclass[review]{elsarticle}
\usepackage[margin=2.0cm]{geometry}
\usepackage{lineno,hyperref}
\usepackage{subfig}
\usepackage{longtable}
\usepackage{caption}
\usepackage{hyperref}
\usepackage{enumitem}
\usepackage{xcolor}
\usepackage{multirow}
\usepackage{subfloat}
\usepackage[linesnumbered,ruled,vlined]{algorithm2e}

\usepackage{float}
\usepackage{amsmath}
\usepackage{setspace}
\usepackage{pdflscape} 
\usepackage{geometry} 

\journal{Journal of \LaTeX\ Templates}
\makeatletter \def\ps@pprintTitle{  \let\@oddhead\@empty  \let\@evenhead\@empty  \def\@oddfoot{\hfill\thepage}  \def\@evenfoot{\thepage\hfill}} \makeatother
\begin{document}

\begin{frontmatter}

\title{usfAD Based Effective Unknown Attack Detection Focused IDS Framework}


\author[1]{Md. Ashraf Uddin}\ead{ashraf.uddin@deakin.edu.au}
\author[1]{Sunil Aryal}\ead{sunil.aryal@deakin.edu.au}  
\author[1]{Mohamed Reda Bouadjenek}
\author[1]{Muna Al-Hawawreh}
\author[2]{Md. Alamin Talukder} \ead{alamin.cse@iubat.edu}

\address[1]{School of Information Technology, Deakin University, Geelong, VIC 3125, Australia}
\address[2]{Department of Computer Science and Engineering, International University of Business Agriculture and Technology, Dhaka, Bangladesh}

\cortext[mycorrespondingauthor]{Corresponding authors: Sunil Aryal and Md Ashraf Uddin}

\begin{abstract}

The rapid expansion of varied network systems, including the Internet of Things (IoT) and Industrial Internet of Things (IIoT), has led to an increasing range of cyber threats. Ensuring robust protection against these threats necessitates the implementation of an effective Intrusion Detection System (IDS). For more than a decade, researchers have delved into  supervised machine learning techniques to develop IDS to classify normal and attack traffic. However, building effective IDS models using supervised learning requires a substantial number of benign and attack samples. To collect a sufficient number of attack samples from real-life scenarios is not possible since cyber attacks occur occasionally. Further, IDS trained and tested on known datasets fails in detecting zero-day or unknown attacks due to the swift evolution of attack patterns. To address this challenge, we put forth two strategies for semi-supervised learning based IDS where training samples of attacks are not required: 1) training a supervised machine learning model using randomly and uniformly dispersed synthetic attack samples; 2) building a One Class Classification (OCC) model that is trained exclusively on benign network traffic. We have implemented both approaches and compared their performances using 10 recent benchmark IDS datasets. Our findings demonstrate that the OCC model based on the state-of-art anomaly detection technique called usfAD significantly outperforms conventional supervised classification and other OCC based techniques when trained and tested considering real-life scenarios, particularly to detect previously unseen attacks.      
\end{abstract}


\end{frontmatter}

\section{Introduction}
\label{Introduction}

Intrusion Detection Systems (IDS) play a crucial role in safeguarding computer networks against cyber attacks \cite{talukder2024machine}. An Intrusion Detection System (IDS) examines network traffic and issues alerts whenever suspicious network or/and system activity is detected \cite{talukder2023dependable}. With the increasing reliance on information network technology, cyber attacks targeting IoT have recently risen significantly. These attacks pose a significant threat not just to IoT but also aggressively target areas crucial to our society, such as national security, corporate data integrity, and public safety. Therefore, it is imperative to develop and deploy IDS for effective detection and prevention of these threats \cite{mahmood2024energy, agate2024blind, aitor2023}.


Many influential research\cite{injadat2020multi}, \cite{gu2021effective}, \cite{kilincer2021machine}, \cite{roy2022lightweight}, \cite{kilincer2022comprehensive}, \cite{naseri2022feature} adopted supervised machine to build IDS where it requires large numbers of both normal and attack instances. These models heavily depend on historical data for training, which might not always include the latest types of cyber attacks. Further, the accuracy of these IDS varies, and they can often yield false negatives (failing to detect actual threats).

We can classify cyberattacks as known and unknown attacks (also called zero-day attacks) in light of IDS's familiarity with the attack during the training phase. Known attacks have specific, identifiable signatures that IDS can recognize based on its training. These attacks are easier to handle as the IDS is already familiar with their characteristics and patterns while training the model. In contrast, unknown attacks present a greater challenge. These are attacks that the IDS has not encountered before, and thus, they lack recognizable patterns. Traditional IDS systems often struggle to identify these unknown attacks because they don't match any expected behavior, profiles, or known attack signatures\cite{fahad2017applying, aghaei2019host, sanchez2021survey, anand2023efficient}. 


IDS that uses supervised Machine Learning (ML) algorithms typically learns patterns of normal and attack categories from training data to classify particular kinds of network traffic instances. Trained IDS can mostly identify testing instances based on their learned patterns \cite{talukder2024mlstl}. ML struggles to correctly identify the class of instances that are not encountered during the training phase. In domains such as network intrusion, and credit card fraud detection, obtaining enough attack instances for training an ML model is challenging due to their scarcity compared to normal traffic/data\cite{talukder2024securing}. In addition, the characteristics of such attacks change swiftly. As a result, in practical scenarios, such models tend to produce a higher number of false negatives (identifying attacks as benign) which is not acceptable in real-world applications. 
The primary concern regarding false negatives is that they enable real threats to remain undiscovered and unaddressed. Such occurrences can result in effective attacks, leading to possible detriment to systems, breaches of data, monetary loss, and injury to an organization's reputation.

\begin{figure}[!htbp]
\centering
\subfloat[F1-score for attack class on NSL-KDD]{\includegraphics[scale=.35]{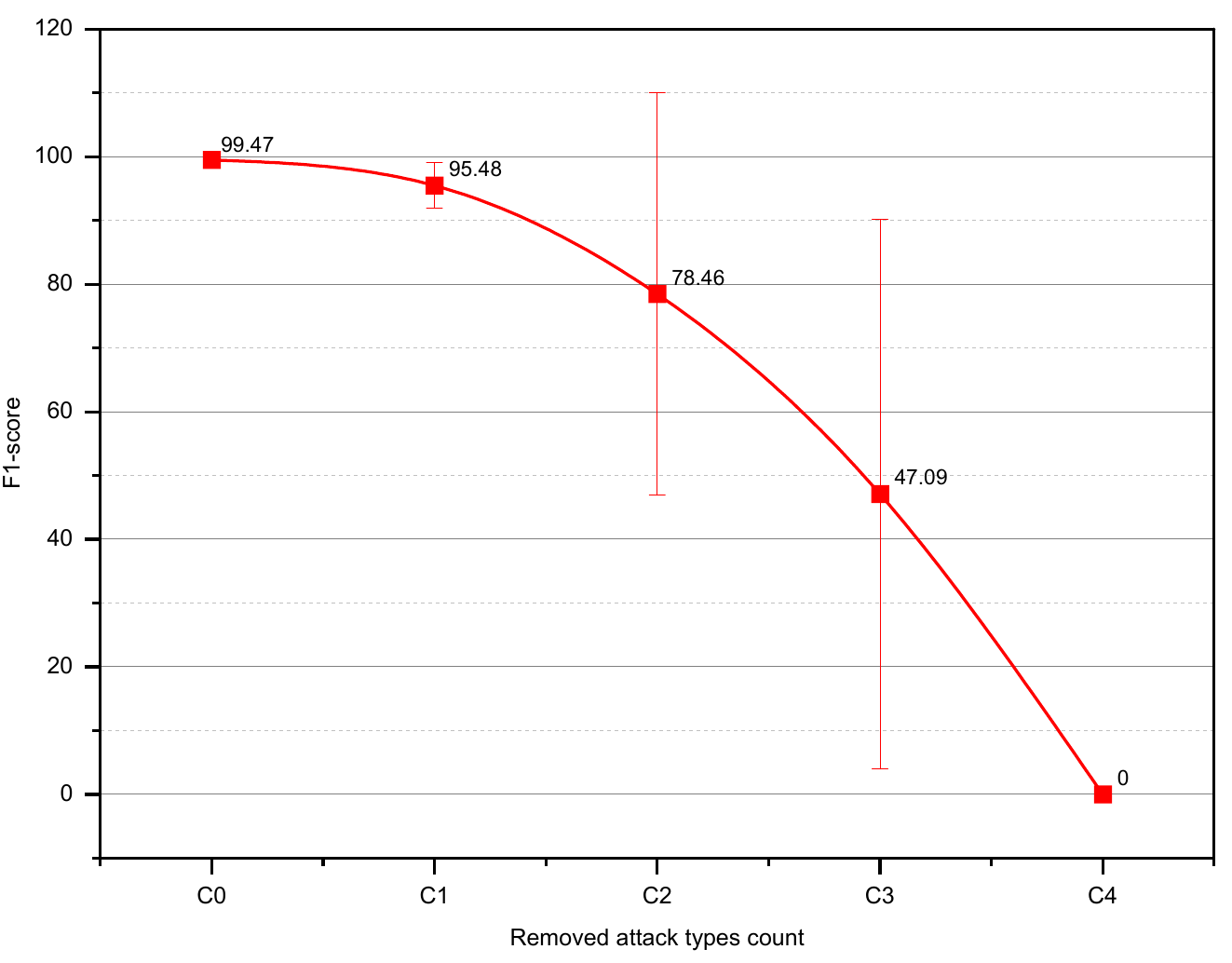}}
\subfloat[F1-score for attack class on UNSW-NB15]{\includegraphics[scale=.35]{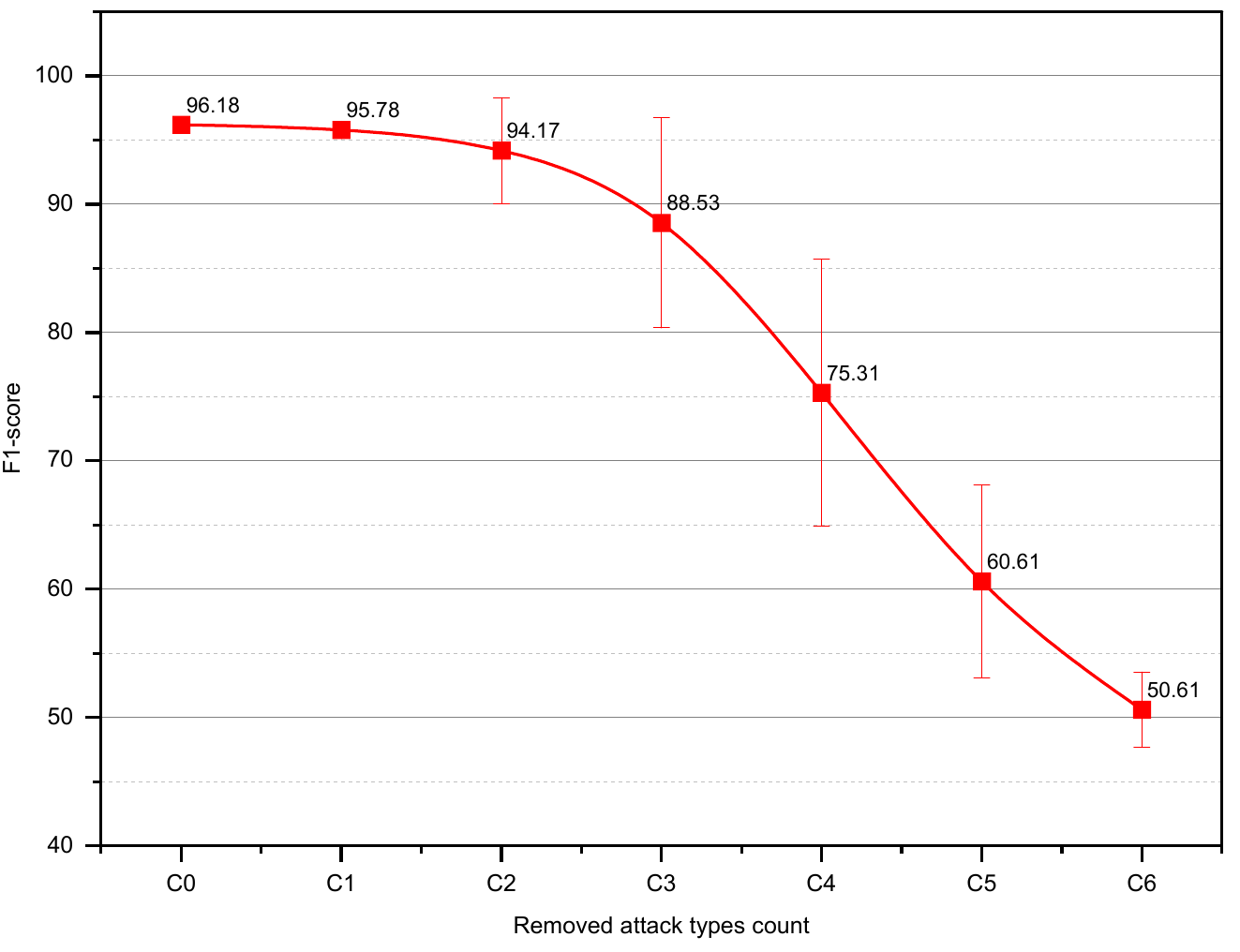}}  
\caption{Impact of removing attack types from training datasets in the RF classifier}
\label{fig:1}
\end{figure}

To demonstrate the above-mentioned limitation of supervised learning-based IDS, we develop and evaluate the capability of the Random Forest (RF), the supervised learning technique that is shown to have superior performance over other counterparts in attack classification\cite{negandhi2019intrusion, liu2021hybrid, wu2022intrusion}, in detecting attacks that are previously unseen during the training phase. We train the RF model to classify attack and normal traffic (i.e., binary classification) by purposefully excluding some attack types from the training set while ensuring all attack types are present in the test set. Note that the task here is to differentiate attack from normal and not to identify attack types correctly. All the experiments are conducted using a 10-fold stratified cross-validation and results in the two widely used IDS benchmark datasets of NSL-KDD and UNSW-NB15  are illustrated in Figure \ref{fig:1}. On the x-axis, labels C0, C1, ..., Cn denote the number of attack types omitted during the training step. C0 signifies that all attack types are incorporated into the training set. C1 means that we sequentially omit each attack category from the training set (first excluding attack category 1, then retaining it while excluding attack category 2, and so on). C2 represents the removal of two attack categories at once, and this pattern continues progressively. The y-axis reflects the average F1-score corresponding to combinations of omissions.  As there are many possible combinations to remove $n$ attack types, we tried all possible combinations and presented the average F1-score and standard deviation. 
Figure \ref{fig:1} indicates that the RF classifier struggles to detect unknown attacks. The observed trend shows that the F1-score for the attack class significantly decreases as samples of more attack classes are excluded while training the RF model.

This observation underscores the ineffectiveness of supervised learning in detecting zero-day or previously unseen attacks. 
Most of the unseen attacks are classified as normal, which can be catastrophic in real-world applications. To tackle this challenge, we explored an alternative approach: training a supervised model using artificially generated data that is uniformly distributed in the feature space and labeled as the "attack" class (simulated attack instances). This strategy allows the model to recognize unseen/unknown attacks without direct training on them. However, our experiment reveals while this approach increases the performance of supervised methods like RF in detecting some unknown attacks, the improvement is not significant enough in high-dimensional real-world datasets to be useful in practical applications. 


Considering supervised models' ineffectiveness in real life situation, we investigate semi-supervised techniques that are trained using available normal data to detect attack data. The semi-supervised learners are closely associated with the subfield of machine learning known as one-class classification (OCC). OCC algorithms aim to model a "normal" class to distinguish unknown data as either normal or attack. 
These techniques are particularly useful for cyberattack detection where training data from the normal class is easily available (because most of the network traffic is benign) while the availability and acquisition of training data from attack classes are limited and challenging\cite{khan2014one}.


In the field of IDS, several researchers\cite{bezerra2019iotds, fahad2017applying, anand2023efficient,dini2022design} investigated several OCC methods, including Local Outlier Factor (LOF)\cite{breunig2000lof}, One-Class SVM (OCSVM)\cite{scholkopf1999support}, Isolation Forest (IF)\cite{liu2008isolation}, and Elliptic Envelope (EE)\cite{rousseeuw1985multivariate}. However, these studies often involve several other steps such as feature engineering and experiments are conducted using a limited set of datasets, which cannot fully represent the effectiveness of one-class classification in this domain. The results presented in these existing studies could be the impact of feature engineering, which may overshadow the true effectiveness of the OCC techniques. Also, empirical evaluations in these studies are limited to a couple of datasets. We require to examine the performance of these models using a wide range of contemporary IDS benchmark datasets. Moreover, more recent  state-of-the-art studies have not yet examined the efficacy of newly developed advanced OCC methods like usfAD (Unsupervised Stochastic Forest based Anomaly Detector) \cite{aryal2018usfad}  and various forms of their ensemble techniques, in the context of network intrusion detection. In \cite{aryal2021usfad}, usfAD has been shown to work well particularly in two cybersecurity datasets. Sunil et al. \cite{aryal2018usfad} developed and applied usfAD to generate scores for detecting outliers. However, we cannot directly adopt usfAD from {aryal2018usfad} to detect network attack samples. Here, we have modified usfAD as OCC by introducing a threshold formula to detect attack categories in IDS.

In this article, our investigation includes a new OCC technique called usfAD and comparing its performance with other popular models. In addition, we construct several ensemble approaches by combining usfAD and other OCC methods to detect network attacks with higher accuracy.We present ensemble approaches, namely Any One, Two, Three, Four, and Five, incorporating our usfAD model along with other state-of-the-art OCC models. The primary objective of these ensemble approaches is to minimize the false-negative rate. The choice of adopting Any One, Two, Three, or Four ensemble approaches depends on the system's resilience against attacks.  By exploring these approaches, we aim to provide a more comprehensive understanding of the effectiveness of one-class classification in the context of IDS.

Our contribution can be summarized as follows. 

\begin{itemize}
  \item We conduct a new experiment to assess the effectiveness of supervised binary classification in detecting unknown or zero-day attacks and investigate the performance of supervised learning to detect unseen attacks by training it using artificially generated attack instances. 
  \item We develop a semi supervised based IDS system for detecting zero-day attacks with higher accuracy. The model includes usfAD and other popular OCC methods and their ensembles. We obtain decision scores from OCC models for each training and testing instance and formulate a customized  threshold to classify network instances as benign or attacks. Our results demonstrate that recently proposed robust outlier detection technique called usfAD achieves higher accuracy with our outlier threshold across the majority of benchmark datasets employed in this study.
 \item We implement and test the model using 10 widely used benchmark IDS datasets to demonstrate its effectiveness in detecting attack instances. We employed 10-runs stratified 80/20 splits to evaluate the model's performance in terms of average accuracy, precision, recall, and F1-score for each modern benchmark IDS dataset. We compare our results with several state-of-the-art works and our findings show that our approach outperforms the existing works.
\end{itemize}


The structure of this paper is as follows: Section \ref{Literature Review} presents a review of related literature. Section \ref{Proposed} details OCC classification architecture and materials used in this study. In Section \ref{RESULTS AND DISCUSSION}, we present the results of our experiments. Finally, Section \ref{Conclusion} summarizes the paper and outlines potential future research directions.

\section{Related works} 
\label{Literature Review}

Most of the existing literature \cite{bezerra2019iotds,anand2023efficient,da2016one,dini2022design}  has primarily focused on detecting attack and normal network instances, which require both normal and attack samples for training. However, in real-life situations, obtaining attack samples is challenging, while normal samples are more readily available. In addition, most existing IDSs built on supervised learning also necessitate correctly labelled data. This makes them unsuitable for real-time use, as they are only able to identify known attacks and are unable to identify novel attack patterns that are not present in their trained dataset. 

However, some IDS models have been built with one class classifier algorithms, which does not require labelled data. These models perform well on some datasets, such NSL-KDD and UNSW-NB15, but they become less effective when used on more recent datasets, like ToN-IoT-Network, CIC-DDoS2019, XIIOTID, and others. The existing models display high false-negative rates, which is harmful for security-related applications where an attack could affect the system as a whole. Consequently, such an approach might not be suitable for real-life scenarios. In this section, we begin by analyzing the most recent OCC methods that are relevant to our study. One-class classifiers possess the capability to train a model without relying on labeled samples of malicious activities. Unlike traditional classifiers that model multiple predefined patterns to evaluate the conformity of new instances, one-class classifiers focus on modeling a single pattern and use it to determine the membership of new instances to that pattern. This approach proves advantageous in the context of IoT devices, which typically exhibit specific behavior characterized by the execution of straightforward tasks while efficiently utilizing computational resources\cite{bezerra2019iotds}.

Extensive research works have been presented in IDS field for detecting attacks using OCC models. For example,  Umer et al.\cite{fahad2017applying} assessed the performance of one-class classification techniques for early detection of malicious flows in a multi-stage flow-based intrusion detection system. The initial stage involved the utilization of minimal flow to classify IP flows as normal or malicious. One-class classification was employed, focusing solely on the malicious class. The performance of the classifier was evaluated using a test dataset comprising both normal and malicious IP flow records. Performance measures such as the Area under the Receiver Operating Characteristic (ROC) curve (AUC) and the F1 score were utilized for result comparison. The findings highlighted the superior accuracy of SVM-based one-class classifiers in detecting malicious IP flows. The v-SVM achieved an AUC of 0.9297 and an F1 score of 0.9114. Based on their experimental results, SVM-based one-classification techniques were deemed suitable for identifying malicious IP flows.

Anand et al.\cite{anand2023efficient} introduced a Machine Learning-based IDS to detect slow rate HTTP/2.0 Denial of Service (DoS) attacks.  They extracted 15 essential features from the datasets. The datasets with minimized number of features were fed into three One-class classifier algorithms: OCSVM, IF, and Minimum Covariant Determinant (MCD). The proposed classifier algorithm outperformed other algorithms in terms of various evaluation measures, including accuracy (0.99), sensitivity (0.99), and specificity (0.99). This highlights the superior performance of their approach in detecting slow rate HTTP/2.0 DoS attacks. An inherent limitation of this study pertains to the training of the one-class classifiers using datasets specific to particular attack types. This constraint arises from the significant variability that characterizes real-world attack scenarios.


The researchers\cite{dini2022design} employed the one-class classifier approach to tackle the challenge of anomaly detection in communication networks. They introduced a novel anomaly detection algorithm that incorporated polynomial interpolation and statistical analysis in its design. This innovative method was applied to well-known datasets widely used in the scientific community, including KDD99, UNSW-NB15, and CSE-CIC-IDS-2018. Additionally, the algorithm was evaluated using a newly available dataset called EDGE-IIOTSET 2022. The study findings showcased that their methodology outperformed traditional one-class classifiers (such as Extreme Learning Machine and Support Vector Machine models) as well as rule-based intrusion detection systems like SNORT in terms of performance.

Wan et al.\cite{wan2017double} introduced a one-class classification anomaly-IDS system method specifically designed for networked control systems, focusing on the dual behavior characteristics. Their approach aimed to provide a clear and understandable solution. By leveraging the unique features of industrial communication, the study aimed to identify and diagnose two specific industrial communication behaviors using a dual one-class classifier approach. The primary objective was to accurately summarize industrial communication behaviors. To accomplish this, the authors proposed the utilization of one-class classifiers, namely OCSVM and RE-KPCA (Reconstruction Error based on Kernel Principal Component Analysis), to detect and classify misbehaviors in industrial communication. They incorporated a weighted mixed kernel function and employed PSO  (Particle Swarm Optimization) parameter optimization to enhance the classification performance. The average accuracy across all three attack types was approximately 83.45\%. Furthermore, the duration of each attack type was similar, with an average duration of approximately 26.11 seconds for all three attack types.

Khraisat et al.\cite{khraisat2020hybrid} developed a Hybrid Intrusion Detection System (HIDS) by combining the C5 decision tree classifier with One Class Support Vector Machine (OC-SVM). The HIDS is designed as a hybrid system that merges the capabilities of Signature-based Intrusion Detection System (SIDS) and Anomaly-based Intrusion Detection System (AIDS). The SIDS algorithm was derived from the C5.0 Decision tree classifier, while the AIDS algorithm was derived from the one-class Support Vector Machine (SVM). The primary objective of this framework was to accurately detect both well-known intrusions and zero-day attacks while minimizing false alarms. To evaluate the effectiveness of the HIDS, the researchers utilized two benchmark datasets: the Network Security Laboratory-Knowledge Discovery in Databases (NSL-KDD) dataset and the Australian Defence Force Academy (ADFA) dataset. The proposed technique successfully integrated the two stages, resulting in outstanding performance with an accuracy rate of 83.24\%. 

We summarized the prior work related to this paper in Table \ref{tab:occliterature}. 


\begin{table}[!htbp]
\centering

\caption{Overview of the related literature.}
\label{tab:occliterature}
\resizebox{\columnwidth}{!}{

\begin{tabular}{|c|l|l|l|}
\hline
Ref & \multicolumn{1}{c|}{Datasets} & \multicolumn{1}{c|}{Models} & \multicolumn{1}{c|}{Remarks} \\ \hline

\cite{fahad2017applying} & CTU-13 dataset & \begin{tabular}[c]{@{}l@{}}Density   Estimation: Simple\\ Gaussian, Mixture of Gaussian,\\ Parzen density estimation,\\ Reconstruction Methods:  AE, \\ SOM, PCA,  Boundary \\ Methods:  v-SVM, SVDD\end{tabular} & v-SVM achieved the highest performance. \\ \hline

\cite{da2016one} & \begin{tabular}[c]{@{}l@{}}SCADA(Supervisory Control\\ and Data Acquisition) systems\end{tabular} & \begin{tabular}[c]{@{}l@{}}SVM,   SVDD(Support Vector \\ Data Description)\end{tabular} & OCSVM was found to perform better. \\ \hline

\cite{anand2023efficient} & Slow rate DoS data & \begin{tabular}[c]{@{}l@{}}SVM, IF, LOF , \\ Elliptic Envelope(EE)\end{tabular} & \begin{tabular}[c]{@{}l@{}}EE achieved the higher accuracy \\ for slow rate HTTP DoS attack.\end{tabular} \\ \hline

\cite{dini2022design} & \begin{tabular}[c]{@{}l@{}}KDD99, UNSW-NB15, CSE-\\ CIC-IDS-2018, EDGE-\\ IIOTSET 2022.\end{tabular} & \begin{tabular}[c]{@{}l@{}}SVM   and Extreme Learning \\ Machine (ELM) with PCA and \\ Polynomial Interpolation\end{tabular} & \begin{tabular}[c]{@{}l@{}}PCA based feature selection was applied. \\ Outcome does not reflect the effectiveness \\ of the SVM and ELM.\end{tabular} \\ \hline

\cite{wan2017double} & SCADA  System & \begin{tabular}[c]{@{}l@{}}SVM-Mixed   Kernel, Guassian \\ Kernel, Polynomial kernel\end{tabular} & \begin{tabular}[c]{@{}l@{}}Mixed kernel based SVM produced higher\\  accuracy.\end{tabular} \\ \hline

\cite{khraisat2020hybrid} & NSL-KDD, ADFA & C5+OCSVM & \begin{tabular}[c]{@{}l@{}}Stacking   ensemble of C5 and One Class \\ SVM were  applied.\end{tabular} \\ \hline

\cite{al2023effective} & Malmem2022 & PCC+OCSVM & \begin{tabular}[c]{@{}l@{}}PCC was used to select the most important \\ features.\end{tabular} \\ \hline

\cite{min2021network} & \begin{tabular}[c]{@{}l@{}}NSL-KDD,   UNSW-NB15, \\ CICIDS 2017\end{tabular} & \begin{tabular}[c]{@{}l@{}}OCSVM,  AE, MemAE, \\ SparseMemAE\end{tabular} & MemAE achieved the higher accuracy. \\ \hline

\cite{mhamdi2020deep} & NSL-KDD & SAE-SVM & \begin{tabular}[c]{@{}l@{}}This approach merged stack\\  auto encoder and one class SVM.\end{tabular} \\ \hline

\cite{nguyen2018nested} & KDD +99 & Nested   OCSVM & Multiple OCSVM was applied. \\ \hline

\cite{arregoces2022network}& UNSW-NB & \begin{tabular}[c]{@{}l@{}}SVM, IF, LOF, EE\end{tabular} & SVM achieved the higher accuracy. \\ \hline

\cite{xu2021improving}& NSL-KDD & AE & \begin{tabular}[c]{@{}l@{}}The work investigated the performance of \\ AE by varying its parameter.\end{tabular} \\ \hline

\cite{alazzam2022lightweight} & \begin{tabular}[c]{@{}l@{}}KDDCUP-99,   UNSW-NB15 , \\ NSL-KDD\end{tabular} & \begin{tabular}[c]{@{}l@{}}OCSVM with Pigeon inspired \\ optimizer\end{tabular} & \begin{tabular}[c]{@{}l@{}}Across all datasets, the author showcased \\ elevated accuracy levels. However, it's \\ noteworthy that achieving such elevated\\ accuracy is not common among \\ researchers, particularly when dealing\\ with the intricate nature of UNSW-NB15.\end{tabular} \\ \hline

\cite{bezerra2019iotds} & BoT-Net & \begin{tabular}[c]{@{}l@{}}(EE, IF,  LOF, and One-class \\ SVM)\end{tabular} & \begin{tabular}[c]{@{}l@{}}Local   Outlier and One Class SVM\\ achieved  the higher accuracy.\end{tabular} \\ \hline

\rotatebox[origin=c]{90}{Proposed OCC} & \begin{tabular}[c]{@{}l@{}}NSL-KDD, UNSW-NB15, \\ ISCXURL2016, Darknet2020, \\ Malmem2022, ToN-IoT-\\ Network,  CIC-DDoS2019, \\ CIC-DoS2017,  XIIOTID, \\ ToN-IoT-Linux\end{tabular} & \begin{tabular}[c]{@{}l@{}}LOF, One Class SVM,  IF, \\ usfAD, AE and VAE\end{tabular} & \begin{tabular}[c]{@{}l@{}}usfAD is found to be performing well \\ across all the datasets\end{tabular} \\ \hline
\end{tabular}
}
\end{table}

In this study, we first aim to investigate methods for employing supervised learning techniques to identify zero-day or previously unknown attacks. However, the challenges of training supervised models in the context of IDS are multiple. Foremost, obtaining an extensive dataset with accurately labeled normal and attack instances is often expensive or unfeasible in real-world scenarios. Although it is possible to acquire accurately labeled attack instances in IDS, these attack instances are infrequent, leading to a skewed class distribution while training a supervised learner. Most existing literature addresses this imbalanced issue by adopting techniques such as under-sampling of normal instances and over-sampling of attack instances. However, this approach tends to generate samples that mirror the distribution of familiar attack patterns. Given that future attacks can manifest in any region of the feature space, diverging from known distributions, the supervised models struggle to generalize over all potential attack patterns when trained on datasets created using conventional balancing techniques. To overcome these challenges, we suggest two primary strategies in this article: i) We can train a supervised algorithm incorporating simulated attack data spread across the entirety of the feature space. This methodology, employed by Aryal et al. \cite{aryal2021ensemble}, focuses on anomaly detection within the data, ii) We can employ techniques that require only the normal or benign data samples, bypassing the need for attack instances entirely. This approach is more suitable in the IDS domain, where normal data samples are more readily accessible.

\section{The Proposed IDS Framework}
\label{Proposed}

In this paper, we discuss two approaches of detecting unknown attacks ( here unknown attacks mean those attacks that are not seen by the model during training phase but appear in the testing datasets). First approach of detecting unknown attack is to train a supervised learner by incorporating dummy attack instances (randomly generated) with original training datasets. The second approach is to adopt a new OCC algorithm called usfAD and different ensembles of the usfAD and other state-of-the-art OCC algorithms to detect unknown attacks.  

In this section, firstly, we train a supervised model using both known normal and attack instances, incorporating uniformly distributed noise data (simulated data) throughout the feature space and labeling them as an "attacks". This approach is adopted from Aryal et al. \cite{aryal2021ensemble}, who introduced the notion of identifying outliers across the local regions of entire feature space. However, their research did not delve into its application in IDS, particularly for classifying novel attack types. Given this, there is an opportunity to explore the idea of integrating noise labeled as "attacks" to identify previously unknown attacks in IDS. In theory, a model learned with such simulated data should possess the capability to identify a broad range of attack instances. As a result, it is anticipated that unknown attacks might be detected as they might align with the distributions learned by the model using the simulated data. 

Secondly, one-class classification (OCC) emerges as a more suitable choice, wherein the model is trained only using normal or benign data. In this paper, we utilize a new OCC method dubbed as usfAD. Below, we first describe the methodology of a supervised model's efficiency in detecting unseen attack instances prior to discussing our OCC based framework.    


\subsection{Methodology for supervised model's Effectiveness in Detecting Unknown Attack}

Prior to explaining the intrusion detection system centered on OCC models, we outline the experimental procedure undertaken to evaluate the potential of supervised learning in identifying unknown attacks without including any noise data or simulated attack data. As a representative of supervised learning, we employ the widely recognized and effective classifier: Random Forest (RF), training and testing it with contemporary benchmark IDS datasets. To assess RF's capability in detecting unknown attacks, we train it using datasets where instances of a specific attack type are removed from the training data while retaining that attack type within the testing data.

\begin{figure}[!htbp]
    \centering
    \includegraphics[scale = .80]{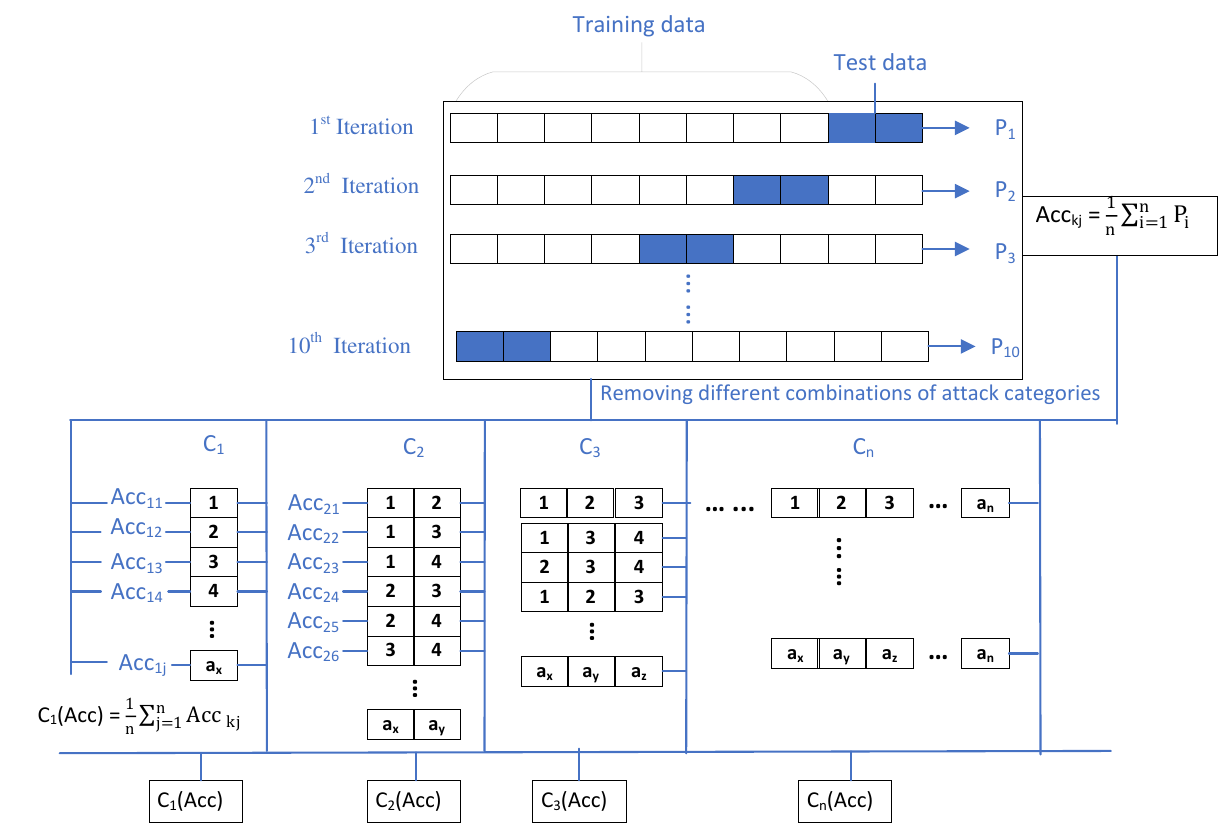}
    \caption{Experiment of Random Forest to detect unknown attacks}
    \label{fig:5}
\end{figure}

Below, we describe the methodological steps of this experiment, aimed at assessing the effectiveness of RF in detecting unknown attacks. We consider a dataset consisting of benign (b) instances and different types of attacks (a1, a2, a3, a4). The process's methodology and algorithm are depicted in Figure \ref{fig:5} and Algorithm \ref{algorithm:1}, respectively.

\begin{itemize}
    \item In this initial step, we create combinations of attacks for omitting from the datasets, such as single attacks {a1}, {a2}, {a3}, and {a4}, two-attack combinations \{a1, a2\}, \{a1, a3\}, \{a1, a4\}, \{a2, a3\}, \{a2, a4\}, \{a3, a4\}, three-attack combinations \{a1, a2, a3\}, \{a1, a3, a4\}, \{a2, a3, a4\}, \{a1, a2, a4\}, four-attack combination \{a1, a2, a3, a4\} and so on.
    \item For omitting each combination, we conduct stratified 10-runs and calculate the average accuracy over these 10 folds. Then, we compute the average accuracy for each combination of varying lengths (one, two, three, or four attacks). For instance, we calculate the average accuracy, precision, recall, and F1-score for one attack combination by summing the 10-fold performance metrics of {a1}, {a2}, {a3}, {a4} and dividing by 4, since there are four distinct combinations with a single attack type. A similar methodology is applied for combinations of two attack types and so forth.
    \item Finally, we represent the accuracy and F1-score values from each length of combination (one, two, three, and four attacks) in a graph (as illustrated in Figure \ref{fig:1}) to visualize the results. This graph provides insights into the performance of the model based on different combinations of attacks.
\end{itemize}

\begin{algorithm}[H]
\caption{Attack Combination for 10 runs}
\label{algorithm:1}
\SetAlgoLined
\SetKwInOut{Input}{Input}
\SetKwInOut{Output}{Output}

\Input{Dataset with instances of benign ($b$) and attacks ($a1, a2, a3, ... ,am$)}
\Output{Accuracy values for different attack combinations}

\For{$k = 1$ to $m$}{
    Generate all $k$-attack combinations $\{C_1, C_2, ..., C_n\}$\;
    \For{$i = 1$ to $n$}{
        Compute the accuracy of combination $C_i$ using stratified 10-runs\;
        Calculate the average accuracy over 10 folds and store it as $Acc_i$\;
    }
    Compute the average accuracy for all $k$-attack combinations: $Avg(Acc_k) = \frac{1}{n} \sum_{i=1}^{n} Acc_i$\;
}
Plot a graph with $x$-axis representing the number of attacks in a combination $(k = 1, 2, 3, ...,m)$, and $y$-axis representing the corresponding average accuracy $Avg(Acc_k)$\;

\end{algorithm}

\subsection{Methodology for Supervised Model's Effectiveness in Detecting Unknown Attack with Simulated Attack Data}

In this case, we incorporate a certain number of simulated attack data (also called noise data) with the original training datasets.  We adopt the similar methodology to validate the efficacy of the RF model in identifying unknown attacks, both with (described in the previous section) and without noise data. In this scenario, we first also form a training datasets after excluding specific combinations of original attack types where removing original attack categories is done based on the previously discussed approach. Next, we incorporate a predetermined quantity of noise instances (dummy or simulated attack data) with the training datasets. Here, we deliberately remove instances of certain attack types from the training datasets and then add simulated attack data so that the model can detect the removed attack types from the training datasets. In real life scenario, we might have enough samples of various types of attack data. So, we expect that a supervised model trained with diverse simulated attack instances might be capable of detecting unknown attacks. The count of dummy attack instances in every combination, where we remove certain attack categories from the training dataset, is defined as:

 $Noise_n = N_n$, where $N_n$ represents the count of regular instances in the original dataset. 
 
This strategy ensures an even distribution between normal and attack data. Given $A_n$ as the count of attack instances and excluding certain combinations denoted as $C_n$, and subsequently adding the noise or simulated instances categorized as attacks, the entire count of instances in the training datasets becomes:

$T_n$ = $N_n+(A_n-C_n)+Noise_n$

For every $C_n$ combination, we incorporate the same set of uniformly distributed, randomly generated noise data.

\subsection{Methodology of usfAD Based IDS}

\begin{figure}[!htbp]
    \centering
    \includegraphics[scale = .70]{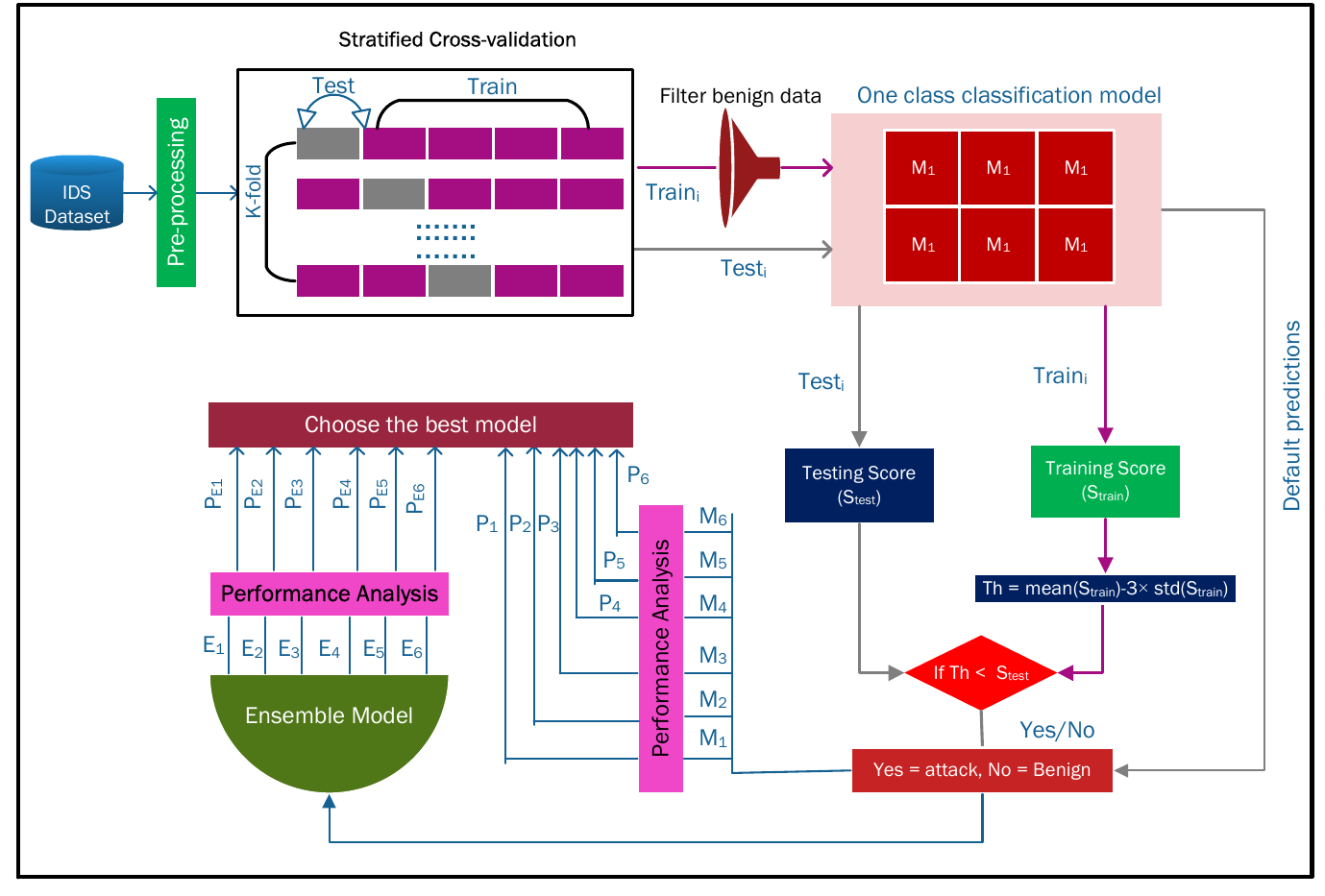}
    \caption{Architecture of One class classification for IDS}
    \label{fig:6}
\end{figure}

In this study, our objective is to develop an OCC based system for the detection of attack instances. The architecture of the model is illustrated in Figure \ref{fig:6}. The benefits of the OCC model are in two folds: this does not necessitate attack samples for training, which overcomes the issue of the limited availability of attack samples in supervised learning. Besides, the dynamic nature of attack characteristics often leads to poor performance of supervised models trained on specific attack types. In addition, supervised learners demand datasets with accurate class labels, which may not be feasible in real-world scenarios. OCC based intrusion detection system addresses these challenges by eliminating the need for explicit dataset labeling and attack samples.

To train the new OCC algorithm called usfAD, we need datasets having only normal or regular network traffic instances. To form a new training dataset devoid of attack instances, we remove all attack instances from the original training dataset. This modified training dataset is used to train usfAD and other standard OCC algorithms including LOF, One-Class SVM, IF, VAE and AE. We also form different ensemble models using the trained OCC models. During the testing phase of these models, any data instance that is not classified as normal is considered as attack instance. The algorithm of the OCC model, we implement in this article is presented in Algorithm \ref{alg:OCCmodel}. Our OCC paradigm is elaborately described below.

\begin{algorithm}[H]
\setstretch{1} 

\caption{Proposed One Class Classifier model}
\label{alg:OCCmodel}
\SetAlgoLined
\SetKwInOut{Input}{Input}
\SetKwInOut{Output}{Output}
\Input{IDS dataset with labeled normal and attack data points, n = number of models, m = number of testing instances, CL = Consensus Level( Any One, Two , Three ensemble and so on}
\Output{Predicted outcomes for individual models and ensemble models}

\For{$f = 1$ to $10$}{
    Split the dataset into  80\% Xtrain[f], Ytrain[f] and 20\% Xtest[f], Ytest[f] \;
    Extract normal points($X^{\prime}_{train}$ and $Y^{\prime}_{train}$) from Xtrain[f] and Ytrain[f]\; 
    \For{$i = 1$ to $n$}{    
       $MT[i]$: Trained model using $X^{\prime}_{train}$ and $Y^{\prime}_{tranin}$ \;
       Generate training data score: Strain[i] = score\_sample($MT_i$,$X^{\prime}_{train}$) \;
       Compute threshold: $TH[i] = \mu (S_{train}) - 3\times \sigma (S_{train})$ \;
       Generate testing score: Stest[i] = score\_sample($MT_i$,$X^{\prime}_{train}$) \;
       \For{$j=1$ to $m$} {
       \eIf{$S_{test}[j]$ $\le$ $TH[i]$ }{
        $predict[i][j] = 1$ }
      {
      $predict[i][j] = 0$
      } 
      }
      }
  \tcp*{Outer loop defines different consensus levels (Any One, Two and so on)}
  \For{$k = 1$ to $CL$}{  
    \For{$j=1$ to $m$}{
      \tcp*{Go through each instance to make a prediction.}
      Initialize attack count $A = 0$\;
      \tcp*{Loop over the predictions from each model}
      \For{$i=1$ to n}{
        \If{$predict[i][j]$==1}{
          Increment $A$\;
        }
      }
      \tcp*{the number of positive predictions is at least k ( consensus level)}
      \eIf{$A \geq k$}{
         $enspredict[k][j] = 1$\;
      }{
             $enspredict[k][j] = 0$\;
        }
      }
    }
  \tcp*{Classification report for individual models}
  $p_f$ = classification\_report($predict_f$, $Ytest_f$])\;
  \tcp*{Classification report for ensemble models}
  $ensp_f$ = classification\_report($enspredict_f$, $Ytest_f$])\;

}

Calculate average performance for 10-folds \;

\end{algorithm}

\begin{itemize}

    \item To prepare our IDS dataset: we require to make sure that the dataset is labeled with normal and attack data points, including different attack types.
    
    \item Train OCC Models: We choose usfAD and multiple popular OCC techniques, such as LOF, IF, OCSVM, VAE and AE $(M_1$, $M_2$, $M_3$,…,$M_n)$.  We train each model using only the normal instance from the IDS dataset. We obtain training score for every instance in the training set and form a threshold score to determine the attack instance. The threshold (TH) is computed as follows: mean (training score)-3× standard deviation (training score) or $TH = \mu-3\times \sigma$ based on the standard 3 sigma rule of statistics. By calculating the threshold based on the mean and standard deviation of the training scores, a boundary is established. Instances with scores above this threshold are considered normal, while instances with scores below the threshold are flagged as potential attacks. For illustrative purposes, Figure \ref{fig:7} displays the decision score derived from the NSL-KDD training datasets, along with its mean and threshold values. As evident from Figure \ref{fig:7}, the green line demarcates a boundary, determining whether scores from testing instances are categorized as an attack or normal.

     \begin{figure}[!htbp]
        \centering
        \includegraphics[scale = .85]{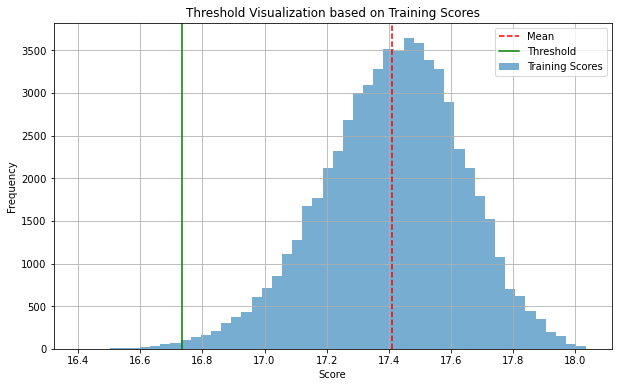}
        \caption{Threshold on the score of training datasets in usfAD model}
        \label{fig:7}
    \end{figure}

    \item Ensemble Model Formation: Subsequently, we construct five different ensemble models by utilizing the prediction results from each individual OCC model. Our ensemble models are: one-model, two-model, three-model, four-model, and five-model ensemble for 10 different datasets. In the case of the one-model ensemble, an attack outcome is determined if any one of the five individual models predicts an attack. Correspondingly, for the two-model ensemble, an attack outcome is derived if at least two of the five individual models predict an attack, and this pattern continues for the other ensemble configurations.

    \item  Performance Determination: Performance for each trained OCC models and their ensemble models is assessed using stratified 10-runs. In this stage, every model determines a traffic instance as attack if the score of the instance is less than the threshold (TH) which is computed based on the scores of the training datasets. We also consider the default prediction outcome of LOF, IF, OCSVM, VAE and AE while the outcome -1 interpreted as 1 or attack and 1 interpreted as 0 or normal.   

\end{itemize}

\subsection{Datasets and Pre-Processing}

We convert categorical values into numerical values, handle missing values by mean imputation, scale the features' values using max-min normalisation. We did not perform other pre-processing techniques such as dealing with outliers, and addressing multicollinearity. Our objective is to evaluate the efficacy of OCC system with minimal data pre-processing, ensuring that the results primarily reflect the intrinsic nature of this approach. We briefly describe the datasets that are used to train our usfAD based model.

\begin{itemize}
    \item NSL-KDD dataset\cite{su2020bat} was designed to overcome the issues with KDD’99 dataset. This updated version of the KDD data set is still regarded as an effective benchmark dataset for researchers to compare different intrusion detection approaches. The NSL-KDD training and testing sets have a balanced quantity of records for benign and attack samples. The shape of the datasets is (148517, 44). Three categorical features (protocol\_type, service, flag) are converted into numerical features using one hot encoding system.   

    \item UNSW-NB15 dataset: The Network Security Research Lab at the University of New South Wales, Australia, built the UNSW-NB15 dataset by capturing network traffic in a realistic setting using a high-speed network sniffer and various tools and techniques such as packet flooding, port scanning, and SQL injection. The original dataset contains 257,673 records and 45 fields. The three categorical features (proto, service, state) are converted into numerical features using one hot encoding method. 

    \item  Canadian Institute for Cybersecurity released CIC-IDS2017 dataset\cite{jazi2017detecting} which is a benchmark dataset for Intrusion Detection System. The dataset includes user behaviour models that are protocol-agnostic through HTTP, HTTPS, FTP, SSH, and email. The dataset consists of 222914,  and 78 features having four classes:  benign samples,  DoS SlowLoris samples,  DoS Slow Httptest samples,  DoS Hulk samples, DoS GoldenEye samples, and Heartbleed samples in the output class label.

    \item CIC-DDoS2019: The Canadian Centre for Cybersecurity at the University of New Brunswick created a dataset of DDoS attacks called CIC-DDoS2019. This data set contains both normal traffic patterns and a wide variety of distributed denial of service (DDoS) assaults, such as UDP flood, HTTP flood, and TCP SYN. The shape of the dataset is(431371, 79) where attack instances are 333540 and benign instances are 97831. 

    \item Malmem2022: Obfuscated malware hides them to avoid detection and elimination using conventional anti-malware software. Malmem 2022\cite{carrier2022detecting} is a simulated obfuscated dataset designed to be realistic as possible to train and test machine learning algorithms to detect obfuscated malware. The dataset is balanced one having level 2 categories:   Spyware, Ransomware, and Trojan Horse.
    
    \item ToN-IoT-Network and ToN-IoT-Linux: ToN-IoT was extracted from a realistic large scale IoT simulated environment at the Cyber Range Lab led by ACCS in 2019. The dataset contains a heterogeneous telemetry IoT services, traffic flows, and logs of operating system. Later, Bro-IDS known as Zeek having 44 features was formed from the original dataset considering the network traffic flows. Label encoding is used to convert its categorical features into numerical features following \cite{moustafa2021new} and \cite{guo2023iot}. These datasets contain IP address. We can treat each unique IP address as a category and perform one-hot encoding. Although this is theoretically possible, it's usually not practical for real-world IDS systems due to the vast number of unique IP addresses, which leads to extremely high-dimensional data.
    
   \item ISCXURL2016: In WWW web, URLs serve as the primary mode of transport and attackers insert malware into users’ computer system through URL. The researchers focus on developing methods for blacklisting malicious URLs. Mamun et al. \cite{mamun2016detecting} formed a modern URL dataset that contains following categories of URLs: benign URLs, spam URLs, phishing URLs, malware URLs and defacement URLs. The shape of the original datasets is (36707, 80). 
   
  \item CIC-Darknet2020: CIC-Darknet dataset has 141530 records with 85 columns features and was labelled in two ways. We apply label encoding to convert its categorical features into numerical values. 

  \item XIIoTID: The XIIoTID dataset\cite{9504604} has an initial shape of (596017, 64).  The dataset has features from  network traffic, system logs, application logs, device's resources (CPU, input/Output, Memory, and others), and commercial Intrusion detection systems' logs (OSSEC and Zeek/Bro). Upon performing one-hot encoding on these features, the total number of features increases to 81.

\end{itemize}






\subsection{Dataset Partitioning and Training}

We do 10 runs of random stratified 80/20 splits to preserve the balanced proportion of each class in each run. This approach provides a more accurate estimate of model performance, particularly when working with imbalanced datasets in which one class is more samples than the other. In this study, we trained and evaluated both the RF and OCC models using the same 10 runs of random stratified 80/20 splits. 


\subsection{Semi-supervised Outlier Detection Algorithms}

In this study, we explore the efficacy of usfAD and other OCC models including LOF, OCSVM, IF, VAE (Variational Autoencoder) and AE (Auto Encoder) to distinguish detect benign and unseen attack instances within IDS datasets. LOF, OCSVM, IF, VAE and AE are well-established outlier detection techniques, and their implementations are sourced from the scikit-learn and PyOD\cite{zhao2019pyod} machine learning library. The implementations of usfAD is obtained from Aryal et al.\cite{aryal2021usfad}. Below, we provide a concise overview of each OCC model.

\begin{itemize}
    \item LOF algorithm\cite{breunig2000lof} is a density-based anomaly detection method. LOF measures the local deviation of a data point with respect to its neighbors. The LOF compares the density of a data point to the densities of its neighbors. If the density of a data point is significantly lower than the densities of its neighbors, the point is likely to be an anomaly.
    
    \item One-Class Support Vector Machines (One-Class SVM)\cite{scholkopf1999support} is an unsupervised machine learning algorithm that is primarily used for novelty detection. OCSVM finds a hyperplane in the feature space that separates the majority of data points from the origin (or a set margin away from the origin) with the largest possible margin. This essentially encompasses the majority of data points in a region, and anything that occurs outside of this region is regarded as an outlier or anomaly.
    \item We adopt usfAD from Aryal et al.\cite{aryal2021usfad} who designed the algorithm based on "Unsupervised Stochastic Forest" (USF) \cite{fernando2017simusf} and Isolation Forest\cite{liu2008isolation}. Unsupervised Stochastic Forest is a variation of unsupervised random forest.  On the other hand, isolation forest, a variant based on the random forest model, offers swift anomaly detection without the dependency on density or distance measures, making it considerably faster than many conventional methods. This usfAD is a robust anomaly score generated technique that does not depend on the scales and units of the data\cite{aryal2021usfad}. 
   \item  Autoencoders (AE)\cite{zhou2017anomaly} trained with normal points can reconstruct these points with minimal error, whereas anomalies or outliers result in higher error. If the reconstruction error surpasses a predefined threshold, the data instance is considered an anomaly or outlier. The AE learns the data distribution of the 'normal' class, and deviations from this distribution (outliers) are more difficult to reconstruct precisely.
   \item Variational Autoencoder (VAE)\cite{an2015variational} is a variation of AE and a generative model that learns to encode and decode data in a way that it can be utilized to detect as anomaly or outlier. If the model is trained primarily on "normal" data, it can reconstruct these normal samples accurately. In contrast, data points that deviate from trained normal data are reconstructed with higher error, thus identifying them outliers.
    
\end{itemize}



\subsection{Experimental Setup and Implementation}

Our experimental study was performed on an Intel Xeon E5-2670 CPU (8 cores, 16 threads), 128GB DDR3 RAM, 2x Nvidia GTX 1080 Ti. Python 3.9 was used to execute our code. The study utilized ten different machine learning models and primarily relied on Pandas and NumPy libraries for data pre-processing. Since the framework was developed using Python, the widely recognized Scikit-learn toolkit was utilized to implement popular outlier methods: LOF, OCSVM, IF and RF classifier. We obtained python code for usfAD from Aryal et al\cite{aryal2021usfad}.

\subsection{Performance Metrics}

In this study, we use accuracy, precision, recall and F1 score that are essential for assessing the performance of an IDS model. However, their significance can vary depending on the system's specific objectives and requirements. Accuracy quantifies the proportion of accurate classifications made by the IDS. However, relying solely on accuracy is not the most suitable performance metric for IDS, as this might not accurately reflect the system's capability to identify attacks, which are a minority class within the dataset. Precision refers to the proportion of genuine positive detection out of all positive detection. High precision is essential in IDS in order to minimise false positives, which can result in false alarms. Recall measures the system's ability to reliably identify all instances of a particular class of attack. Low recall suggests that the system is missing some attacks, which can pose a significant security risk.

The F1 score is a combination of precision and recall that quantifies the proportion of true positive identification relative to the total number of positive instances in the dataset. F1-score is a valuable metric for IDS because this considers both false positives and false negatives and provides a balanced score between precision and recall. The accuracy, precision, recall and F1-score are calculated as follows.

\[
\text{Accuracy} = \frac{\text{TP + TN}}{\text{TP + TN + FP + FN}} \times 100
\]

\[
\text{Precision} = \frac{\text{TP}}{\text{TP+FP}} \times 100
\]

\[
\text{Recall} = \frac{\text{TP}}{\text{TP+FN}} \times 100
\]

\[
\text{F1-score} = 2\times \frac{\text{Precision $\times$ Recall}}{\text{Precision + Recall}} \times 100
\]

where TP = true positive, TN = true negative, FP = false positive, and FN = false negative.

\section{Results and Discussion}
\label{RESULTS AND DISCUSSION}

In this section, we evaluate the effectiveness of the OCC model and RF classifier while detecting unknown attacks. We present the average accuracy, precision, recall, and F1-score for OCC and RF models.

\subsection{Performance of Supervised Learning to Detect Unknown Attacks}

In our experiment, we aimed to assess the effectiveness of RF model in detecting unknown attacks in the context of IDS. In this section, we present the performance of a RF trained with uniformly distributed synthetic noise data labeled as attacks and without noise data. First, we examine the effect of adding noise data in identifying attacks on a synthetic dataset with two features to see it visually. Second, we investigate the effectiveness of this approach on real IDS datasets with the two most important features. In this case, we consider real IDS datasets with two features because real datasets are intricate and adhere to varied distributions and when employing a random function to produce noise data with a substantial number of features, the noise does not span the entire 0 to 1 spectrum. Consequently, the presence of this noise data does not significantly influence the detection of unseen attacks. Our findings show that when the model is trained using datasets with a complete set of features, the performance remains the same for both RF with noise and without noise.

\subsubsection{ RF Model Trained on Synthetic Data with Noise}

\begin{figure}[!htbp]
    \centering
    \includegraphics[scale = .85]{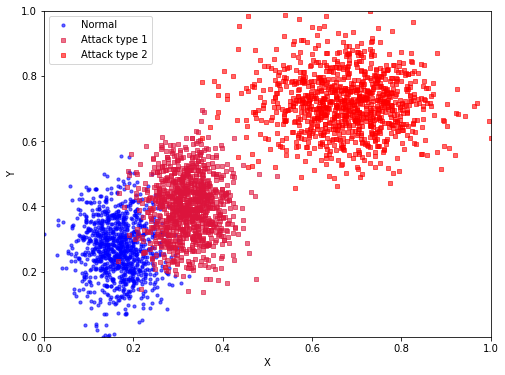}
    \caption{Simulated datasets for RF's efficiency}
    \label{fig:2}
\end{figure}

Our first solution to detect unknown attacks is to train a supervised model using random uniformly distributed data (labeled as an attack) along with the original dataset. To evaluate the impact of adding external noise to the results, we crafted synthetic datasets with a Gaussian distribution, visualized in Figure \ref{fig:2}. Here, blue, crimson, and red clusters represent benign data, attack type 1, and attack type 2, respectively. We consider the synthetic datasets with two features, allowing for visualization of the model's decision boundary and its predictive outcomes, illustrating its adaptability when encountering noise and its ability to identify unknown attacks.

\begin{figure}[!htbp]
    \centering
    \includegraphics[scale = .80]{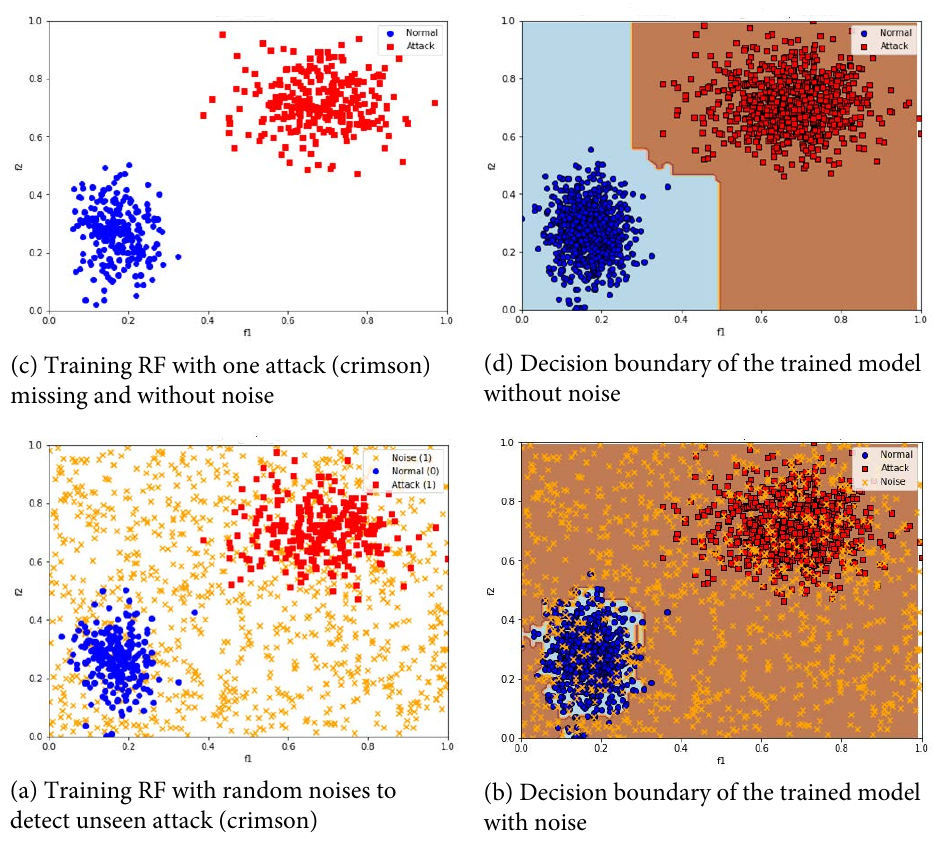}
    \caption{Training RF including noise instances and only normal instances }
    \label{fig:3}
\end{figure}

\begin{figure}[!htbp]
    \centering
    \includegraphics[scale = .75]{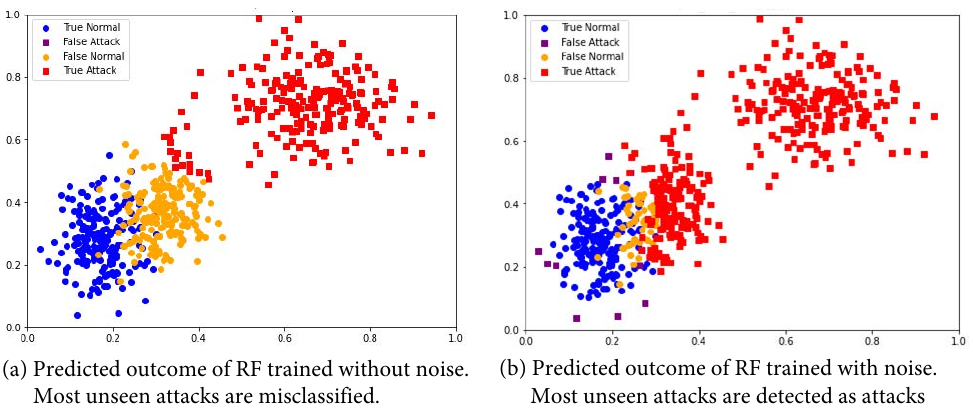}
    \caption{Predicted outcome of RF with noises and normal data only}
    \label{fig:4}
\end{figure}

In this experiment, we present the synthetic dataset with one normal cluster and two types of attacks in Figure \ref{fig:2}. To show the capability of RF model to detect unknown attacks, we simulate unknown attack type case by removing the middle attack type (crimson)(as shown in Figure \ref{fig:2}) during the training stage as shown in Figure \ref{fig:3} (a). Figure \ref{fig:3} (b) present the decision boundary of the trained model with missing attack type 1 and without noise. Next, we add noise during the training stage to make up for the unknown attacks.
Figure \ref{fig:3} (c) display the training datasets containing normal, attack type 2 and noise data (depicted as blue, red and orange points) and Figure \ref{fig:3} (d) shows the decision boundary of the trained RF model with noise.

\begin{figure}[!htbp]
    \centering
    \includegraphics[scale = .75]{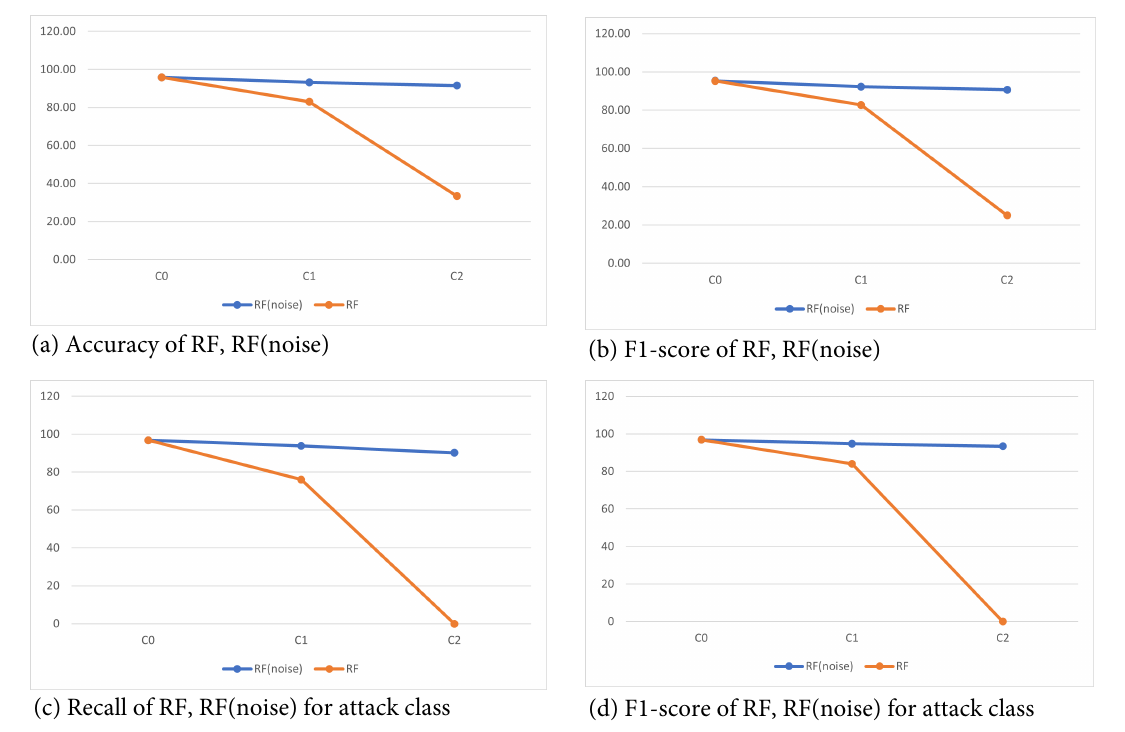}
    \caption{Performance on synthetic datasets}
    \label{fig:41}
\end{figure}

In both case, we make attack type 1 unknown to the model. Notably, the decision boundary shown in Figure \ref{fig:3} (b) is more constricted than the one in Figure \ref{fig:3} (d), attributable to the introduction of noise data. In Figure \ref{fig:4} (a), we observe that all instances of attack type 2 and most of attack type 1 are correctly identified because of the noise data, even though the model was not specifically trained on attack type 2. Within this figure, a red square signifies a correctly classified attack, while a orange circle indicates a misclassification. Conversely, Figure \ref{fig:4} (b) portrays that a model trained with normal data and attack type 2 struggles to correctly identify attack type 1 as this attack type was not present in the training datasets. We can see that most of the attack type 2 are identified as normal (orange circle). Figure \ref{fig:41} (a) and (b) display the average macro accuracy and F1-score of the synthetic datasets for two scenarios: RF (noise), and standard RF. Meanwhile, the recall and F1-score focused on the attack class are depicted in Figure \ref{fig:41} (c) and (d). As observed from the plots in Figure \ref{fig:41}, the RF trained with noise exhibits a superior ability in identifying unseen attacks compared to the standard RF. 


\subsubsection{RF Model Trained with Benchmark IDS Data with Noise}

\begin{figure}[!htbp]
    \centering
    \includegraphics[scale = .70]{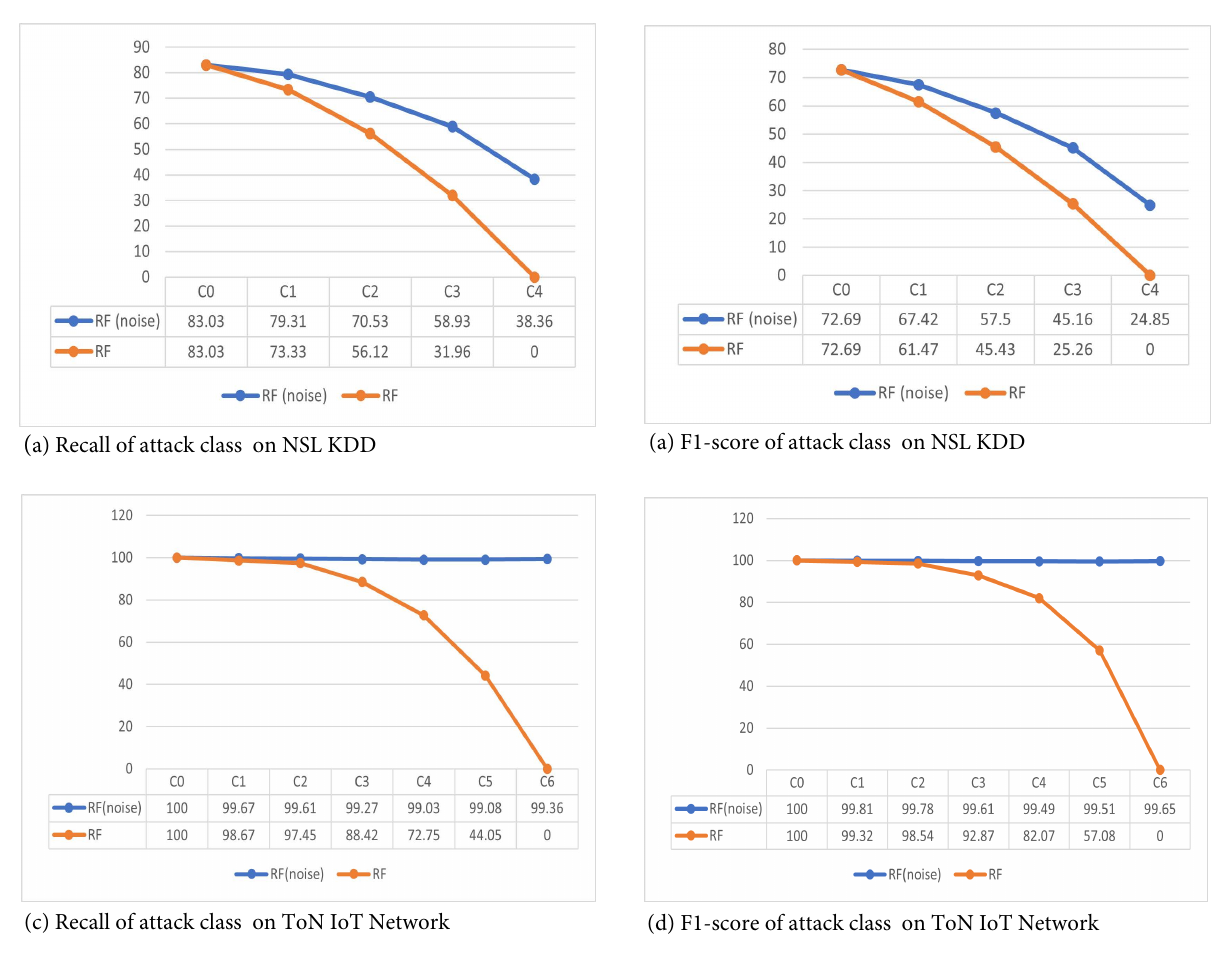}
    \caption{RF's performance in detecting unknown attacks}
    \label{fig:8}
\end{figure}

The results showcased here are based on benchmark IDS datasets that incorporate only two features selected using Random Forest \cite{li2020building, disha2022performance}. We have chosen to illustrate the effect of noise on benchmark IDS datasets by focusing on just two features. This decision is based on the observation that using Random Forest (RF) on high-dimensional feature sets alongside noise leads to unstable results. Furthermore, introducing noise across a large number of dimensions incurs a substantial computational expense. This makes the process of incorporating noise into the training of a supervised model for the detection of unseen attacks impractical. Below, we present the outcome of training a RF model by adding noise data labelled as attack with benchmark IDS datasets. We consider the IDS datasets with the most important two features. The experimental results are depicted in Figure \ref{fig:8}, which illustrates a gradual drop in accuracy as different types of attacks are removed from the training dataset, while the testing data contains all attack types. The methodology section explains the process of removing various attack categories from the training data and computing the accuracy to depict the graph. We conduct this experiment for 10 different benchmark IDS datasets including NSL-KDD, UNSW-NB15, ISCXURL2016, CIC-DoS2017, CIC-DDoS2019, CIC-Darknet2020, CIC-Malmem2022, ToN-IoT-Network, ToN-IoT-Linux, and XIIOTID datasets. 

In Figure \ref{fig:8}, we show outcome of RF models after inclusion of noise in identifying previously unknown attack samples. In order to empower the RF model to identify unknown attacks, we adopt a strategy that involves training the model with the original datasets combined with uniformly distributed random datasets. These additional datasets share the same number of features as the originals. In this approach, the added data instances are labeled as attacks, aligning with the goal of enhancing the model's capability to recognize zero-day attacks. The count of noise records matches the number of normal instances. To evaluate the impact of introducing these random datasets on the RF model's ability to detect unknown attack samples, we conducted training and testing using various training datasets. Each training dataset involves the removal of specific quantities of attack instances, but importantly, the introduced noise data (present in 80\% of the training data) is retained in every case. The testing dataset, on the other hand, exclusively comprises original attack instances and do not include noise samples.

The graph depicted in Figure \ref{fig:8} showcases the outcomes of a series of experiments involving the removal of varying numbers of attack class instances during the model training process. On the x-axis, different scenarios are presented, each representing the removal of a specific count of attack type instances from the training dataset. The corresponding recall and F1-score for attack class for each scenario is depicted on the y-axis.

Figure \ref{fig:8} presents recall and F1-score of the RF model trained with and without noise data across scenarios where varying numbers of attack types are excluded. Both the RF model trained with noise and the one without noise exhibit identical performance, with a recall and F1-score of 83.03\%, 72.69\% and 100\%, 100\% for NSL-KDD, and ToN-IoT Network with the most important two features. This suggests that, when all attack types are present in the training data, introducing noise does not have a discernible positive or negative effect on the model's recall and F1-score. For C3, the RF model with noise data maintains better performance (C3: 58.93\%, 45.16\% and 99.27\%, 99.61\% for NSL-KDD, and UNSW-NB15) compared to the RF model without noise (C3: 31.96\%, 25.26\% and 88.42\%, 92.87\%). As we move from C3 to C4, and C6 the gap in performance widens significantly. The trend suggests that as more attack types are excluded, the RF model without noise struggles more, highlighting the advantage of using noise in training for better generalization. At C4, the RF model with noise data manages to achieve a recall and F1-score of 38.36\%, 24.85\% for NSL KDD and 99.36\%, 99.65\% for ToN-IoT-Network datasets at C6. However, the RF model without noise fails entirely, resulting in 0\% for both recall and F1-score on both datasets. This means the standard RF model couldn't correctly classify any instances under this configuration, emphasizing the importance of the noise data when dealing with unseen attack types.

The RF model trained with noise data consistently shows superior performance compared to the one without noise, particularly when multiple attack types are removed. Introducing noise during training appears to bolster the model's ability to recognize unfamiliar or less common attack variants. Such adaptability is vital in real-world contexts, where unpredictable attack types may arise.

\begin{figure}[!htbp]
    \centering
    \includegraphics[scale = .70]{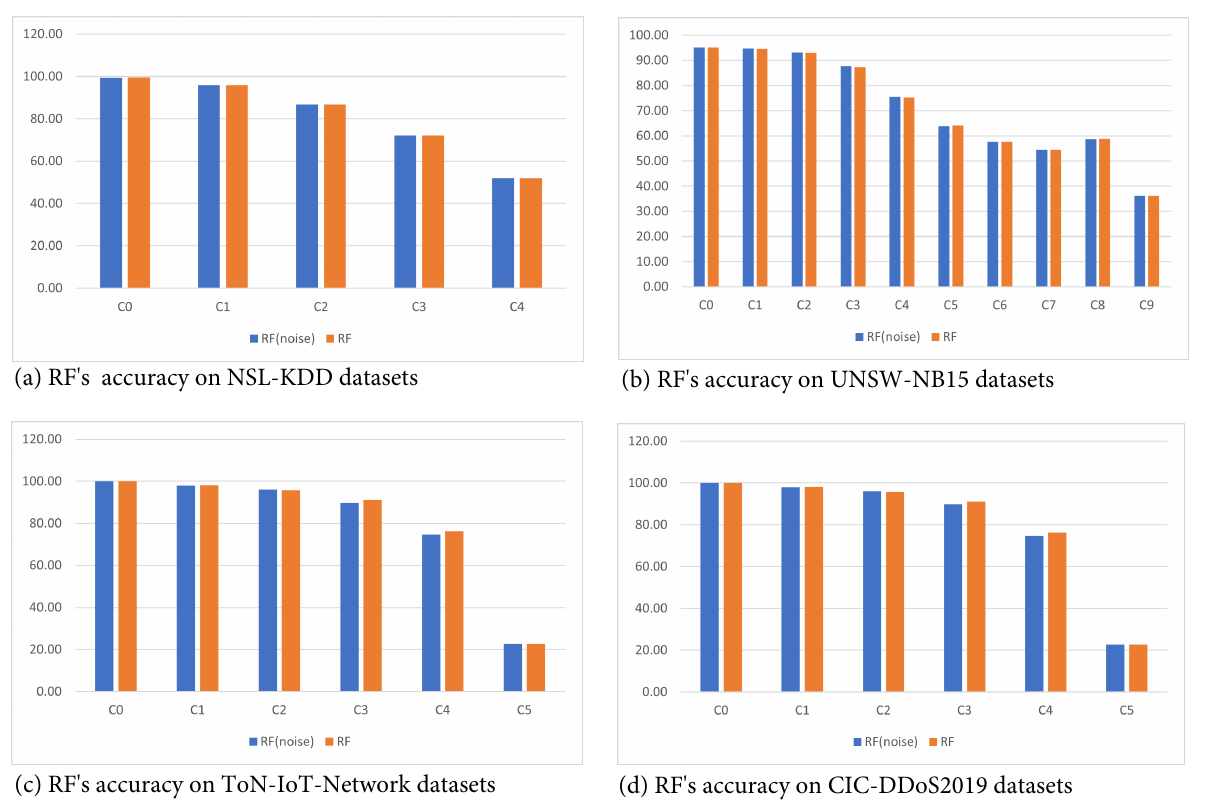}
    \caption{RF's accuracy in detecting unseen attacks with full featured datasets}
    \label{fig:81}
\end{figure}

Nonetheless, our experiments indicate that while introducing noise can enhance the effectiveness of supervised methods such as Random Forest (RF) in identifying certain novel attacks, the degree of improvement on high-dimensional real-world datasets is not substantial enough to warrant practical application. As depicted in Figure \ref{fig:81}, there is essentially no discernible difference between the performance of the RF model with noise and without it. This outcome may stem from the fact that the randomly generated noise does not cover the full range of the binary space. Moreover, the generation of noise in a high-dimensional space entails considerable computational resources.

As a result, we are now focusing on exploring one-class classifiers or outlier detection methods as an alternative approach. Our goal is to detect network intrusions and compare the performance of this approach with the conventional binary Random Forest classifier in a real-life scenario.

\subsection{Performance of Semi Supervised Learning or OCC}

In this section, we discuss the outcome of semi supervised or OCC methods and their ensemble approaches. Table \ref{tab:outlierperformance1} displays the accuracy and F1-score results for various semi-supervised learning algorithms, including LOF, IF, OCSVM, VAE, AE, and usfAD, along with different ensemble approaches. In Table \ref{tab:outlierperformance1}, we present the accuracy evaluation conducted on 10 distinct datasets: NSL-KDD, UNSW-NB15, ISCXURL2016, Malmem2022 and CIC-DDoS2019, TON-IOT-Network, CIC-Darknet2020, CIC-DoS2017, XIIOTID, and ToN-IoT-Linux datasets. For the first three datasets—NSL-KDD, UNSW-NB15, and ISCXURL2016 dataset, we can notice that VAE and AE exhibited almost the same performance. For this reason, we chose to exclude it from evaluations on other large datasets since VAE demands substantial memory demands for large IDS datasets.

\begin{table}[!htbp]
\centering
\caption{ Accuracy of outlier methods and their ensemble approaches}
\label{tab:outlierperformance1}
\begin{tabular}{|lllllllllll|}
\hline
\multicolumn{1}{|l|}{\multirow{2}{*}{Models}} & \multicolumn{10}{c|}{IDS Benchmark Datasets} \\ \cline{2-11} 
\multicolumn{1}{|l|}{} & \multicolumn{1}{l|}{D1} & \multicolumn{1}{l|}{D2} & \multicolumn{1}{l|}{D3} & \multicolumn{1}{l|}{D4} & \multicolumn{1}{c|}{D5} & \multicolumn{1}{c|}{D6} & \multicolumn{1}{c|}{D7} & \multicolumn{1}{c|}{D8} & \multicolumn{1}{c|}{D9} & \multicolumn{1}{c|}{D10} \\ \hline
\multicolumn{1}{|l|}{LOF} & \multicolumn{1}{l|}{87.50} & \multicolumn{1}{l|}{80.87} & \multicolumn{1}{l|}{80.95} & \multicolumn{1}{l|}{88.00} & \multicolumn{1}{l|}{86.60} & \multicolumn{1}{l|}{98.19} & \multicolumn{1}{l|}{\textbf{93.10}} & \multicolumn{1}{l|}{83.90} & \multicolumn{1}{l|}{78.49} & 96.70 \\ \hline
\multicolumn{1}{|l|}{VAE} & \multicolumn{1}{l|}{89.03} & \multicolumn{1}{l|}{54.65} & \multicolumn{1}{l|}{73.26} & \multicolumn{1}{l|}{---} & \multicolumn{1}{l|}{---} & \multicolumn{1}{l|}{---} & \multicolumn{1}{l|}{---} & \multicolumn{1}{l|}{---} & \multicolumn{1}{l|}{---} & --- \\ \hline
\multicolumn{1}{|l|}{AE} & \multicolumn{1}{l|}{91.53} & \multicolumn{1}{l|}{56.26} & \multicolumn{1}{l|}{74.03} & \multicolumn{1}{l|}{94.98} & \multicolumn{1}{l|}{79.18} & \multicolumn{1}{l|}{60.81} & \multicolumn{1}{l|}{76.91} & \multicolumn{1}{l|}{86.48} & \multicolumn{1}{l|}{82.42} & 67.34 \\ \hline
\multicolumn{1}{|l|}{OCSVM} & \multicolumn{1}{l|}{73.54} & \multicolumn{1}{l|}{73.91} & \multicolumn{1}{l|}{81.27} & \multicolumn{1}{l|}{75.03} & \multicolumn{1}{l|}{84.81} & \multicolumn{1}{l|}{67.54} & \multicolumn{1}{l|}{50.08} & \multicolumn{1}{l|}{52.30} & \multicolumn{1}{l|}{68.42} & 45.74 \\ \hline
\multicolumn{1}{|l|}{IF} & \multicolumn{1}{l|}{90.28} & \multicolumn{1}{l|}{56.95} & \multicolumn{1}{l|}{65.09} & \multicolumn{1}{l|}{90.36} & \multicolumn{1}{l|}{70.00} & \multicolumn{1}{l|}{63.21} & \multicolumn{1}{l|}{78.30} & \multicolumn{1}{l|}{88.94} & \multicolumn{1}{l|}{71.05} & 66.58 \\ \hline
\multicolumn{1}{|l|}{usfAD} & \multicolumn{1}{l|}{\textbf{95.92}} & \multicolumn{1}{l|}{\textbf{82.15}} & \multicolumn{1}{l|}{92.38} & \multicolumn{1}{l|}{94.65} & \multicolumn{1}{l|}{\textbf{98.69}} & \multicolumn{1}{l|}{\textbf{99.43}} & \multicolumn{1}{l|}{91.65} & \multicolumn{1}{l|}{\textbf{97.04}} & \multicolumn{1}{l|}{\textbf{93.52}} & \textbf{97.94} \\ \hline
\multicolumn{1}{|l|}{} & \multicolumn{10}{c|}{Ensemble-Any One, Two, Three, Four, and Five} \\ \hline
\multicolumn{1}{|l|}{Ensemble-1} & \multicolumn{1}{l|}{71.16} & \multicolumn{1}{l|}{75.01} & \multicolumn{1}{l|}{85.74} & \multicolumn{1}{l|}{70.00} & \multicolumn{1}{l|}{85.50} & \multicolumn{1}{l|}{66.24} & \multicolumn{1}{l|}{55.43} & \multicolumn{1}{l|}{44.97} & \multicolumn{1}{l|}{69.87} & 64.94 \\ \hline
\multicolumn{1}{|l|}{Ensemble-2} & \multicolumn{1}{l|}{89.97} & \multicolumn{1}{l|}{79.81} & \multicolumn{1}{l|}{88.92} & \multicolumn{1}{l|}{86.60} & \multicolumn{1}{l|}{94.46} & \multicolumn{1}{l|}{91.38} & \multicolumn{1}{l|}{84.03} & \multicolumn{1}{l|}{83.50} & \multicolumn{1}{l|}{91.04} & 90.81 \\ \hline
\multicolumn{1}{|l|}{Ensemble-3} & \multicolumn{1}{l|}{94.27} & \multicolumn{1}{l|}{78.71} & \multicolumn{1}{l|}{85.37} & \multicolumn{1}{l|}{93.47} & \multicolumn{1}{l|}{93.72} & \multicolumn{1}{l|}{96.34} & \multicolumn{1}{l|}{83.43} & \multicolumn{1}{l|}{91.50} & \multicolumn{1}{l|}{91.74} & 76.90 \\ \hline
\multicolumn{1}{|l|}{Ensemble-4} & \multicolumn{1}{l|}{94.54} & \multicolumn{1}{l|}{60.82} & \multicolumn{1}{l|}{76.82} & \multicolumn{1}{l|}{\textbf{97.98}} & \multicolumn{1}{l|}{82.09} & \multicolumn{1}{l|}{68.92} & \multicolumn{1}{l|}{83.56} & \multicolumn{1}{l|}{94.15} & \multicolumn{1}{l|}{80.13} & 72.90 \\ \hline
\multicolumn{1}{|l|}{Ensemble-5} & \multicolumn{1}{l|}{88.83} & \multicolumn{1}{l|}{55.79} & \multicolumn{1}{l|}{72.23} & \multicolumn{1}{l|}{94.96} & \multicolumn{1}{l|}{63.50} & \multicolumn{1}{l|}{66.30} & \multicolumn{1}{l|}{83.59} & \multicolumn{1}{l|}{94.51} & \multicolumn{1}{l|}{61.12} & 68.76 \\ \hline
\multicolumn{11}{|l|}{\begin{tabular}[c]{@{}l@{}}D1 = NSL-KDD, D2 = UNSW-NB15, D3 = ISCXURL2016, D4 = Malmem2020,  D5 = CIC-DDoS2019,\\ D6 = ToN-IoT-Network, D7 =  Darknet2020, D8 = CIC-DoS2017, D9 = XIIOTID, ToN-IoT-Linux\end{tabular}} \\ \hline
\end{tabular}
\end{table}

The usfAD appears to be a standout performer, achieving high accuracy and F1-scores even on these datasets, both individually and in ensemble settings. For example, the usfAD achieved the highest accuracy 94.96\%, 80.23\%, 92.38\% , and 98.69\% respectively for NSL-KDD, UNSW-NB15, ISCXURL2016, and CIC-DDoS2019. Particularly commendable performances by usfAD are evident in the ToN-IoT-Network and ToN-IoT-Linux datasets, where it achieved accuracy of 99.43\% and 97.94\%, respectively. However, considering individual model, AE always shows better accuracy for the Malmem2022 dataset, and the Ensemble-Any Four strategy outperforms other outlier methods, securing an accuracy of 97.98\% for this dataset. This underlines the efficacy of the ensemble strategy introduced in this study.

For the UNSW NB15 dataset, LOF achieves an accuracy and F1-score of 80.87\% respectively, closely rivaling usfAD. When considering ensemble configurations, the Ensemble-Any Four approach parallels usfAD's performance on the NSL-KDD dataset. On the UNSW-NB15 and ISCXURL2016 datasets, Ensemble-Any Two outperforms other outlier detection methods, including LOF, IF, OCSVM, VAE, and AE. It's noteworthy that VAE and AE yield nearly identical results. Due to this similarity, and given VAE's intensive memory requirements, we opted to exclude VAE for other datasets including Malmem2022 and CIC-DDoS2019.

Table \ref{tab:outlierperformance2} presents the macro average F1-score results of the semi-supervised learning for 10 different IDS datasets. usfAD consistently delivers strong results across all datasets, underscoring its suitability for handling IDS datasets in terms of macro average F1-score. 

Particularly commendable performances by usfAD are evident in the ToN-IoT-Network and ToN-IoT-Linux datasets, where it achieved F1-scores of 99.37\% and 97.65\% respectively. LOF also demonstrates a strong performance on the ToN-IoT-Linux dataset but has varying results on the other two datasets in terms of F1-score. Such results are on par with supervised algorithms like RF.

\begin{table}[!htbp]
\centering
\caption{Macro average F1-score of outlier methods and their ensemble approaches}
\label{tab:outlierperformance2}
\begin{tabular}{|lllllllllll|}
\hline
\multicolumn{1}{|c|}{Models} & \multicolumn{1}{l|}{D1} & \multicolumn{1}{l|}{D2} & \multicolumn{1}{l|}{D3} & \multicolumn{1}{l|}{D4} & \multicolumn{1}{c|}{D5} & \multicolumn{1}{c|}{D6} & \multicolumn{1}{c|}{D7} & \multicolumn{1}{c|}{D8} & \multicolumn{1}{c|}{D9} & \multicolumn{1}{c|}{D10} \\ \hline
\multicolumn{1}{|l|}{LOF} & \multicolumn{1}{l|}{87.48} & \multicolumn{1}{l|}{80.61} & \multicolumn{1}{l|}{76.58} & \multicolumn{1}{l|}{87.99} & \multicolumn{1}{l|}{88.29} & \multicolumn{1}{l|}{98.03} & \multicolumn{1}{l|}{\textbf{88.37}} & \multicolumn{1}{l|}{64.03} & \multicolumn{1}{l|}{75.88} & 96.35 \\ \hline
\multicolumn{1}{|l|}{VAE} & \multicolumn{1}{l|}{89.01} & \multicolumn{1}{l|}{53.75} & \multicolumn{1}{l|}{69.46} & \multicolumn{1}{l|}{---} & \multicolumn{1}{l|}{---} & \multicolumn{1}{l|}{---} & \multicolumn{1}{l|}{---} & \multicolumn{1}{l|}{---} & \multicolumn{1}{l|}{---} & --- \\ \hline
\multicolumn{1}{|l|}{AE} & \multicolumn{1}{l|}{91.52} & \multicolumn{1}{l|}{55.93} & \multicolumn{1}{l|}{70.18} & \multicolumn{1}{l|}{94.97} & \multicolumn{1}{l|}{75.58} & \multicolumn{1}{l|}{42.55} & \multicolumn{1}{l|}{51.73} & \multicolumn{1}{l|}{58.49} & \multicolumn{1}{l|}{81.75} & 54.86 \\ \hline
\multicolumn{1}{|l|}{OCSVM} & \multicolumn{1}{l|}{72.24} & \multicolumn{1}{l|}{69.56} & \multicolumn{1}{l|}{70.82} & \multicolumn{1}{l|}{73.37} & \multicolumn{1}{l|}{75.38} & \multicolumn{1}{l|}{67.53} & \multicolumn{1}{l|}{44.07} & \multicolumn{1}{l|}{42.56} & \multicolumn{1}{l|}{67.97} & 43.31 \\ \hline
\multicolumn{1}{|l|}{IF} & \multicolumn{1}{l|}{90.22} & \multicolumn{1}{l|}{56.37} & \multicolumn{1}{l|}{62.53} & \multicolumn{1}{l|}{90.26} & \multicolumn{1}{l|}{54.03} & \multicolumn{1}{l|}{46.21} & \multicolumn{1}{l|}{51.94} & \multicolumn{1}{l|}{58.72} & \multicolumn{1}{l|}{67.13} & 45.74 \\ \hline
\multicolumn{1}{|l|}{usfAD} & \multicolumn{1}{l|}{\textbf{95.91}} & \multicolumn{1}{l|}{\textbf{81.84}} & \multicolumn{1}{l|}{\textbf{87.61}} & \multicolumn{1}{l|}{94.64} & \multicolumn{1}{l|}{\textbf{98.04}} & \multicolumn{1}{l|}{\textbf{99.37}} & \multicolumn{1}{l|}{84.06} & \multicolumn{1}{l|}{\textbf{88.49}} & \multicolumn{1}{l|}{\textbf{93.32}} & \textbf{97.65} \\ \hline
\multicolumn{1}{|l|}{} & \multicolumn{1}{l|}{} & \multicolumn{1}{l|}{} & \multicolumn{1}{l|}{} & \multicolumn{1}{l|}{} & \multicolumn{1}{l|}{} & \multicolumn{1}{l|}{} & \multicolumn{1}{l|}{} & \multicolumn{1}{l|}{} & \multicolumn{1}{l|}{} &  \\ \hline
\multicolumn{1}{|l|}{Ensemble-1} & \multicolumn{1}{l|}{69.22} & \multicolumn{1}{l|}{70.04} & \multicolumn{1}{l|}{72.86} & \multicolumn{1}{l|}{67.04} & \multicolumn{1}{l|}{72.53} & \multicolumn{1}{l|}{66.20} & \multicolumn{1}{l|}{53.32} & \multicolumn{1}{l|}{38.44} & \multicolumn{1}{l|}{69.23} & 64.94 \\ \hline
\multicolumn{1}{|l|}{Ensemble-2} & \multicolumn{1}{l|}{89.95} & \multicolumn{1}{l|}{79.19} & \multicolumn{1}{l|}{83.03} & \multicolumn{1}{l|}{86.36} & \multicolumn{1}{l|}{91.31} & \multicolumn{1}{l|}{90.96} & \multicolumn{1}{l|}{76.03} & \multicolumn{1}{l|}{65.79} & \multicolumn{1}{l|}{90.98} & 90.08 \\ \hline
\multicolumn{1}{|l|}{Ensemble-3} & \multicolumn{1}{l|}{94.27} & \multicolumn{1}{l|}{78.46} & \multicolumn{1}{l|}{80.83} & \multicolumn{1}{l|}{93.44} & \multicolumn{1}{l|}{91.16} & \multicolumn{1}{l|}{96.06} & \multicolumn{1}{l|}{66.67} & \multicolumn{1}{l|}{73.05} & \multicolumn{1}{l|}{91.55} & 68.62 \\ \hline
\multicolumn{1}{|l|}{Ensemble-4} & \multicolumn{1}{l|}{94.50} & \multicolumn{1}{l|}{60.39} & \multicolumn{1}{l|}{72.93} & \multicolumn{1}{l|}{ \textbf{97.98}} & \multicolumn{1}{l|}{82.10} & \multicolumn{1}{l|}{51.28} & \multicolumn{1}{l|}{55.75} & \multicolumn{1}{l|}{69.97} & \multicolumn{1}{l|}{77.92} & 57.86 \\ \hline
\multicolumn{1}{|l|}{Ensemble-5} & \multicolumn{1}{l|}{88.58} & \multicolumn{1}{l|}{54.59} & \multicolumn{1}{l|}{69.10} & \multicolumn{1}{l|}{94.94} & \multicolumn{1}{l|}{50.73} & \multicolumn{1}{l|}{43.38} & \multicolumn{1}{l|}{51.03} & \multicolumn{1}{l|}{62.93} & \multicolumn{1}{l|}{46.94} & 46.60 \\ \hline
\multicolumn{11}{|c|}{\begin{tabular}[c]{@{}c@{}}D1 = NSL-KDD, D2 = UNSW-NB15, D3 = ISCXURL2016, D4 =   Malmem2020,  D5 = CIC-DDoS2019,\\ D6 = ToN-IoT-Network, D7 = Darknet2020, D8 = CIC-DoS2017, D9 = XIIOTID, D10 = ToN-IoT-Linux\end{tabular}} \\ \hline
\end{tabular}
\end{table}


Table \ref{tab:attackclassperformance1}, and \ref{tab:attackclassperformance2} present the performance metrics of various outlier detection models and their ensemble approaches in terms of average precision and recall of 10-fold stratified cross validation for the attack class only, across 10 distinct datasets.  Recall, also known as sensitivity or true positive rate, is a crucial metric in the context of outlier detection for security-related tasks. High recall is desirable in scenarios where missing any attack instance is considered highly detrimental, as is often the case in cybersecurity. Our goal in this work is to accurately identify attack class while taking into account the importance of such classification in cybersecurity applications. In cybersecurity, misclassifying an attack as normal poses a greater threat to the system than incorrectly labeling a normal instance as an attack. Because of this, we focus on the average recall and precision of the attack class for the outliers/OCC techniques. 
 

\begin{table}[!htbp]
\centering
\caption{Precision of outlier methods and their ensemble approaches for attack class}
\label{tab:attackclassperformance1}
\begin{tabular}{|lllllllllll|}
\hline
\multicolumn{1}{|c|}{Models} & \multicolumn{1}{l|}{D1} & \multicolumn{1}{l|}{D2} & \multicolumn{1}{l|}{D3} & \multicolumn{1}{l|}{D4} & \multicolumn{1}{c|}{D5} & \multicolumn{1}{c|}{D6} & \multicolumn{1}{c|}{D7} & \multicolumn{1}{c|}{D8} & \multicolumn{1}{c|}{D9} & \multicolumn{1}{c|}{D10} \\ \hline
\multicolumn{1}{|l|}{LOF} & \multicolumn{1}{l|}{87.13} & \multicolumn{1}{l|}{96.86} & \multicolumn{1}{l|}{96.39} & \multicolumn{1}{l|}{86.21} & \multicolumn{1}{l|}{95.08} & \multicolumn{1}{l|}{95.08} & \multicolumn{1}{l|}{76.95} & \multicolumn{1}{l|}{24.75} & \multicolumn{1}{l|}{96.32} & 92.66 \\ \hline
\multicolumn{1}{|l|}{VAE} & \multicolumn{1}{l|}{89.11} & \multicolumn{1}{l|}{85.73} & \multicolumn{1}{l|}{96.12} & \multicolumn{1}{l|}{---} & \multicolumn{1}{l|}{---} & \multicolumn{1}{l|}{---} & \multicolumn{1}{l|}{---} & \multicolumn{1}{l|}{---} & \multicolumn{1}{l|}{---} & --- \\ \hline
\multicolumn{1}{|l|}{AE} & \multicolumn{1}{l|}{89.65} & \multicolumn{1}{l|}{86.80} & \multicolumn{1}{l|}{96.23} & \multicolumn{1}{l|}{90.96} & \multicolumn{1}{l|}{25.48} & \multicolumn{1}{l|}{25.48} & \multicolumn{1}{l|}{22.10} & \multicolumn{1}{l|}{18.89} & \multicolumn{1}{l|}{84.82} & 52.47 \\ \hline
\multicolumn{1}{|l|}{OCSVM} & \multicolumn{1}{l|}{64.74} & \multicolumn{1}{l|}{75.59} & \multicolumn{1}{l|}{87.07} & \multicolumn{1}{l|}{66.69} & \multicolumn{1}{l|}{51.84} & \multicolumn{1}{l|}{51.84} & \multicolumn{1}{l|}{17.29} & \multicolumn{1}{l|}{10.60} & \multicolumn{1}{l|}{58.74} & 27.27 \\ \hline
\multicolumn{1}{|l|}{IF} & \multicolumn{1}{l|}{93.82} & \multicolumn{1}{l|}{92.28} & \multicolumn{1}{l|}{96.40} & \multicolumn{1}{l|}{83.86} & \multicolumn{1}{l|}{39.40} & \multicolumn{1}{l|}{39.40} & \multicolumn{1}{l|}{24.45} & \multicolumn{1}{l|}{20.64} & \multicolumn{1}{l|}{82.99} & 49.57 \\ \hline
\multicolumn{1}{|l|}{usfAD} & \multicolumn{1}{l|}{96.47} & \multicolumn{1}{l|}{97.04} & \multicolumn{1}{l|}{92.78} & \multicolumn{1}{l|}{90.34} & \multicolumn{1}{l|}{98.39} & \multicolumn{1}{l|}{98.39} & \multicolumn{1}{l|}{77.69} & \multicolumn{1}{l|}{72.50} & \multicolumn{1}{l|}{96.90} & 99.29 \\ \hline
\multicolumn{1}{|l|}{} & \multicolumn{1}{l|}{\textbf{}} & \multicolumn{1}{l|}{\textbf{}} & \multicolumn{1}{l|}{\textbf{}} & \multicolumn{1}{l|}{\textbf{}} & \multicolumn{1}{l|}{\textbf{}} & \multicolumn{1}{l|}{} & \multicolumn{1}{l|}{} & \multicolumn{1}{l|}{} & \multicolumn{1}{l|}{} &  \\ \hline
\multicolumn{1}{|l|}{Ensemble-1} & \multicolumn{1}{l|}{62.53} & \multicolumn{1}{l|}{75.33} & \multicolumn{1}{l|}{85.83} & \multicolumn{1}{l|}{62.50} & \multicolumn{1}{l|}{50.85} & \multicolumn{1}{l|}{50.85} & \multicolumn{1}{l|}{27.76} & \multicolumn{1}{l|}{10.14} & \multicolumn{1}{l|}{59.45} & 48.77 \\ \hline
\multicolumn{1}{|l|}{Ensemble-2} & \multicolumn{1}{l|}{84.03} & \multicolumn{1}{l|}{90.97} & \multicolumn{1}{l|}{92.27} & \multicolumn{1}{l|}{78.88} & \multicolumn{1}{l|}{80.20} & \multicolumn{1}{l|}{80.20} & \multicolumn{1}{l|}{52.01} & \multicolumn{1}{l|}{26.59} & \multicolumn{1}{l|}{85.83} & 80.62 \\ \hline
\multicolumn{1}{|l|}{Ensemble-3} & \multicolumn{1}{l|}{93.74} & \multicolumn{1}{l|}{95.41} & \multicolumn{1}{l|}{95.91} & \multicolumn{1}{l|}{88.45} & \multicolumn{1}{l|}{90.52} & \multicolumn{1}{l|}{90.52} & \multicolumn{1}{l|}{52.36} & \multicolumn{1}{l|}{40.10} & \multicolumn{1}{l|}{92.18} & 83.69 \\ \hline
\multicolumn{1}{|l|}{Ensemble-4} & \multicolumn{1}{l|}{98.39} & \multicolumn{1}{l|}{98.00} & \multicolumn{1}{l|}{97.04} & \multicolumn{1}{l|}{96.15} & \multicolumn{1}{l|}{89.14} & \multicolumn{1}{l|}{89.14} & \multicolumn{1}{l|}{60.38} & \multicolumn{1}{l|}{56.18} & \multicolumn{1}{l|}{97.17} & 95.49 \\ \hline
\multicolumn{1}{|l|}{Ensemble-5} & \multicolumn{1}{l|}{99.64} & \multicolumn{1}{l|}{99.56} & \multicolumn{1}{l|}{98.14} & \multicolumn{1}{l|}{99.51} & \multicolumn{1}{l|}{91.16} & \multicolumn{1}{l|}{91.16} & \multicolumn{1}{l|}{79.67} & \multicolumn{1}{l|}{77.26} & \multicolumn{1}{l|}{97.07} & 98.41 \\ \hline
\multicolumn{11}{|l|}{\begin{tabular}[c]{@{}l@{}}D1 = NSL-KDD, D2 = UNSW-NB15, D3 = ISCXURL2016, D4 = Malmem2020, D5 = CIC-DDoS2019, \\ D6 = ToN-IoT-Network, D7 = Darknet2020, D8 = CIC-DoS2017, D9 = XIIOTID, D10 = ToN-IoT-Linux\end{tabular}} \\ \hline
\end{tabular}
\end{table}

Across CIC-DDoS2019 and ToN-IoT-Network datasets, LOF displays relatively high recall values, indicating its effectiveness in identifying attacks. Its precision is generally good, suggesting a balanced performance. Auto Encoder: Auto Encoder's precision and recall vary greatly across datasets. While it demonstrates high precision in Malmem2022, its recall is generally low, implying it struggles to capture all positive instances, especially in CIC-DDoS2019 and TON-IOT-Network.  

Table \ref{tab:attackclassperformance2} shows that the usfAD achieves high recall across all datasets except Darknet2020 and ToN-IoT-Linux among individual models. For example, on the XIIOTID dataset, the usfAD achieves a precision of 96.9\% and a recall of 87.92\%. Similarly, on the ToN-IoT-Linux dataset, it reaches an impressive 99.29\% precision and 94.49\% recall. For Darknet2020 and ToN-IoT-Linux, LOF model shows relatively higher recall among the individual models.

\begin{table}[!htbp]
\centering
\caption{Recall of outlier methods and their ensemble approaches for attack class}
\label{tab:attackclassperformance2}
\begin{tabular}{|lllllllllll|}
\hline
\multicolumn{1}{|c|}{Models} & \multicolumn{1}{l|}{D1} & \multicolumn{1}{l|}{D2} & \multicolumn{1}{l|}{D3} & \multicolumn{1}{l|}{D4} & \multicolumn{1}{c|}{D5} & \multicolumn{1}{c|}{D6} & \multicolumn{1}{c|}{D7} & \multicolumn{1}{c|}{D8} & \multicolumn{1}{c|}{D9} & \multicolumn{1}{c|}{D10} \\ \hline
\multicolumn{1}{|l|}{LOF} & \multicolumn{1}{l|}{86.86} & \multicolumn{1}{l|}{72.42} & \multicolumn{1}{l|}{78.78} & \multicolumn{1}{l|}{90.46} & \multicolumn{1}{l|}{100} & \multicolumn{1}{l|}{100.00} & \multicolumn{1}{l|}{\textbf{85.39}} & \multicolumn{1}{l|}{75.64} & \multicolumn{1}{l|}{52.57} & \textbf{97.89} \\ \hline
\multicolumn{1}{|l|}{VAE} & \multicolumn{1}{l|}{87.95} & \multicolumn{1}{l|}{34.14} & \multicolumn{1}{l|}{68.84} & \multicolumn{1}{l|}{---} & \multicolumn{1}{l|}{---} & \multicolumn{1}{l|}{---} & \multicolumn{1}{l|}{---} & \multicolumn{1}{l|}{---} & \multicolumn{1}{l|}{---} & --- \\ \hline
\multicolumn{1}{|l|}{AE} & \multicolumn{1}{l|}{93.15} & \multicolumn{1}{l|}{37.22} & \multicolumn{1}{l|}{69.78} & \multicolumn{1}{l|}{99.88} & \multicolumn{1}{l|}{6.35} & \multicolumn{1}{l|}{6.35} & \multicolumn{1}{l|}{13.64} & \multicolumn{1}{l|}{34.49} & \multicolumn{1}{l|}{72.60} & 22.13 \\ \hline
\multicolumn{1}{|l|}{OCSVM} & \multicolumn{1}{l|}{98.86} & \multicolumn{1}{l|}{87.39} & \multicolumn{1}{l|}{89.53} & \multicolumn{1}{l|}{100.00} & \multicolumn{1}{l|}{100.00} & \multicolumn{1}{l|}{100.00} & \multicolumn{1}{l|}{50.37} & \multicolumn{1}{l|}{87.92} & \multicolumn{1}{l|}{92.19} & 37.51 \\ \hline
\multicolumn{1}{|l|}{IF} & \multicolumn{1}{l|}{85.45} & \multicolumn{1}{l|}{35.62} & \multicolumn{1}{l|}{57.87} & \multicolumn{1}{l|}{99.98} & \multicolumn{1}{l|}{10.03} & \multicolumn{1}{l|}{10.03} & \multicolumn{1}{l|}{12.36} & \multicolumn{1}{l|}{27.16} & \multicolumn{1}{l|}{42.12} & 6.91 \\ \hline
\multicolumn{1}{|l|}{usfAD} & \multicolumn{1}{l|}{\textbf{94.99}} & \multicolumn{1}{l|}{74.33} & \multicolumn{1}{l|}{\textbf{97.95}} & \multicolumn{1}{l|}{\textbf{100.00}} & \multicolumn{1}{l|}{\textbf{100.00}} & \multicolumn{1}{l|}{\textbf{100.00}} & \multicolumn{1}{l|}{70.40} & \multicolumn{1}{l|}{\textbf{85.79}} & \multicolumn{1}{l|}{\textbf{87.92}} & 94.49 \\ \hline
\multicolumn{1}{|l|}{} & \multicolumn{1}{l|}{\textbf{}} & \multicolumn{1}{l|}{\textbf{}} & \multicolumn{1}{l|}{\textbf{}} & \multicolumn{1}{l|}{\textbf{}} & \multicolumn{1}{l|}{\textbf{}} & \multicolumn{1}{l|}{} & \multicolumn{1}{l|}{} & \multicolumn{1}{l|}{} & \multicolumn{1}{l|}{} &  \\ \hline
\multicolumn{1}{|l|}{Ensemble-1} & \multicolumn{1}{l|}{100.00} & \multicolumn{1}{l|}{90.53} & \multicolumn{1}{l|}{98.09} & \multicolumn{1}{l|}{100.00} & \multicolumn{1}{l|}{100.00} & \multicolumn{1}{l|}{100.00} & \multicolumn{1}{l|}{99.50} & \multicolumn{1}{l|}{97.83} & \multicolumn{1}{l|}{96.83} & 99.03 \\ \hline
\multicolumn{1}{|l|}{Ensemble-2} & \multicolumn{1}{l|}{97.72} & \multicolumn{1}{l|}{75.95} & \multicolumn{1}{l|}{93.80} & \multicolumn{1}{l|}{100.00} & \multicolumn{1}{l|}{100.00} & \multicolumn{1}{l|}{100.00} & \multicolumn{1}{l|}{77.61} & \multicolumn{1}{l|}{91.15} & \multicolumn{1}{l|}{95.12} & 95.40 \\ \hline
\multicolumn{1}{|l|}{Ensemble-3} & \multicolumn{1}{l|}{94.40} & \multicolumn{1}{l|}{70.05} & \multicolumn{1}{l|}{85.07} & \multicolumn{1}{l|}{99.99} & \multicolumn{1}{l|}{100.00} & \multicolumn{1}{l|}{100.00} & \multicolumn{1}{l|}{36.62} & \multicolumn{1}{l|}{69.20} & \multicolumn{1}{l|}{88.52} & 38.25 \\ \hline
\multicolumn{1}{|l|}{Ensemble-4} & \multicolumn{1}{l|}{90.12} & \multicolumn{1}{l|}{39.50} & \multicolumn{1}{l|}{72.81} & \multicolumn{1}{l|}{99.96} & \multicolumn{1}{l|}{12.54} & \multicolumn{1}{l|}{12.54} & \multicolumn{1}{l|}{12.48} & \multicolumn{1}{l|}{34.89} & \multicolumn{1}{l|}{55.97} & 19.74 \\ \hline
\multicolumn{1}{|l|}{Ensemble-5} & \multicolumn{1}{l|}{77.06} & \multicolumn{1}{l|}{30.96} & \multicolumn{1}{l|}{66.01} & \multicolumn{1}{l|}{90.36} & \multicolumn{1}{l|}{3.84} & \multicolumn{1}{l|}{3.84} & \multicolumn{1}{l|}{5.97} & \multicolumn{1}{l|}{17.92} & \multicolumn{1}{l|}{10.97} & 6.51 \\ \hline
\multicolumn{11}{|c|}{\begin{tabular}[c]{@{}c@{}}D1 = NSL-KDD, D2 =   UNSW-NB15, D3 = ISCXURL2016, D4 = Malmem2020, D5 = CIC-DDoS2019, \\ D6 = ToN-IoT-Network, D7 = Darknet2020, D8 = CIC-DoS2017, D9 = XIIOTID, D10 = ToN-IoT-Linux\end{tabular}} \\ \hline
\end{tabular}
\end{table}

Our examination of the recall metrics reveals that although the usfAD system exhibits commendable performance across various datasets, its ability to detect the majority of attacks is limited in specific cases, such as the USNW-NB15, Darknet2020, and XIIOTOD datasets. In practical scenarios, achieving a high rate of attack detection is crucial. To address this shortfall, one could consider employing various ensemble strategies using unsupervised models. As shown in Table \ref{tab:attackclassperformance2}, the Ensemble-1 approach, which flags an attack if any individual model reports one, achieves superior recall rates at the cost of precision in datasets like NSL-KDD, UNSW-NB15, and Darknet2020. Depending on the operational requirements, system administrators can decide which aspect—recall or precision—is more critical for their situation. By ensembling multiple models, one can enhance the recall rate of attacks, albeit at the cost of an increased number of false positives (where normal instances are incorrectly identified as attacks).


\subsubsection{Performance of Ensemble Approaches}

\begin{figure}[!htbp]
    \centering
    \includegraphics[scale = .75]{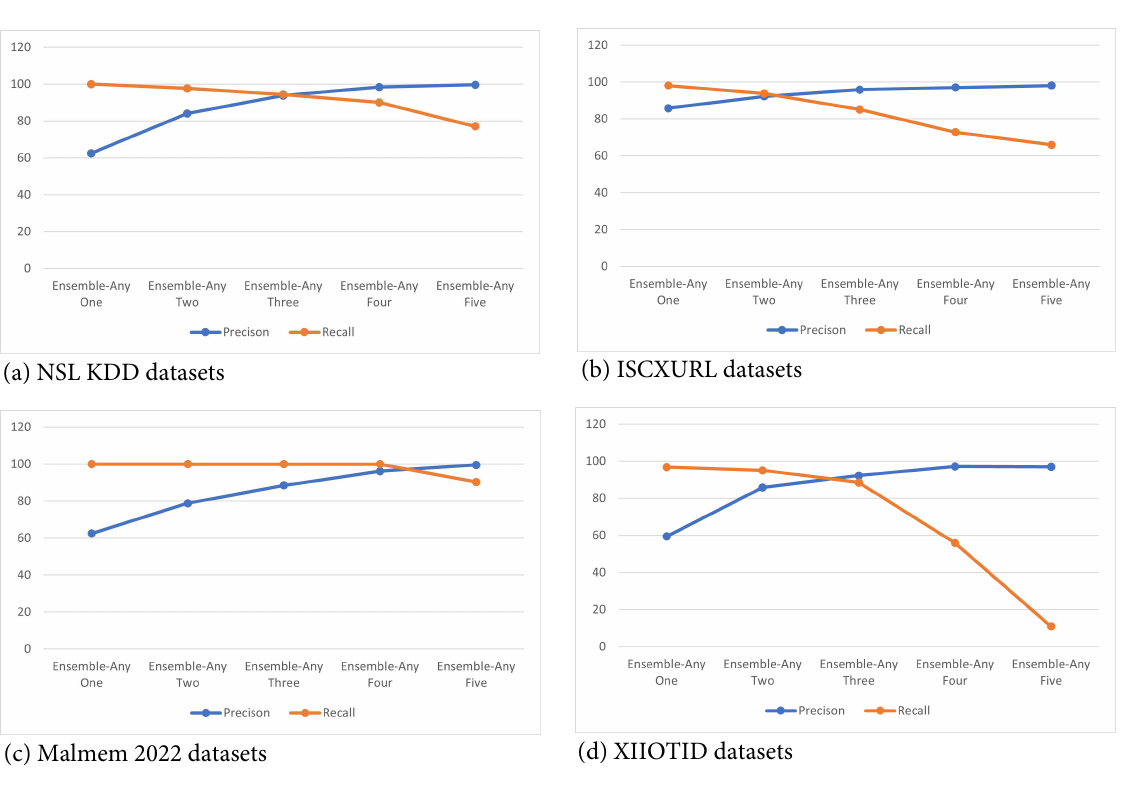}
    \caption{Performance of different ensemble approaches for attack class}
    \label{fig:9}
\end{figure}

We present the performance of different ensemble approaches in Figure \ref{fig:9}. As observed in Figure \ref{fig:9}, some ensemble approaches have high recall but lower precision, while others exhibit high precision but lower recall. The choice between these models depends on the specific application's priorities. For instance, in security applications where missing any attack instance is critical, models with higher recall might be preferred.



A trade-off between precision and recall needs to be considered. In situations where avoiding false negatives (missed attacks) is critical, models or ensembles with higher recall, such as usfAD and certain ensemble approaches, might be preferred. However, in scenarios where precision is more important, trade-offs must be carefully evaluated.


\subsection{RF vs. usfAD: Detecting Unknown Attacks}

\begin{figure}[!htbp]
    \centering
    \includegraphics[scale = .65]{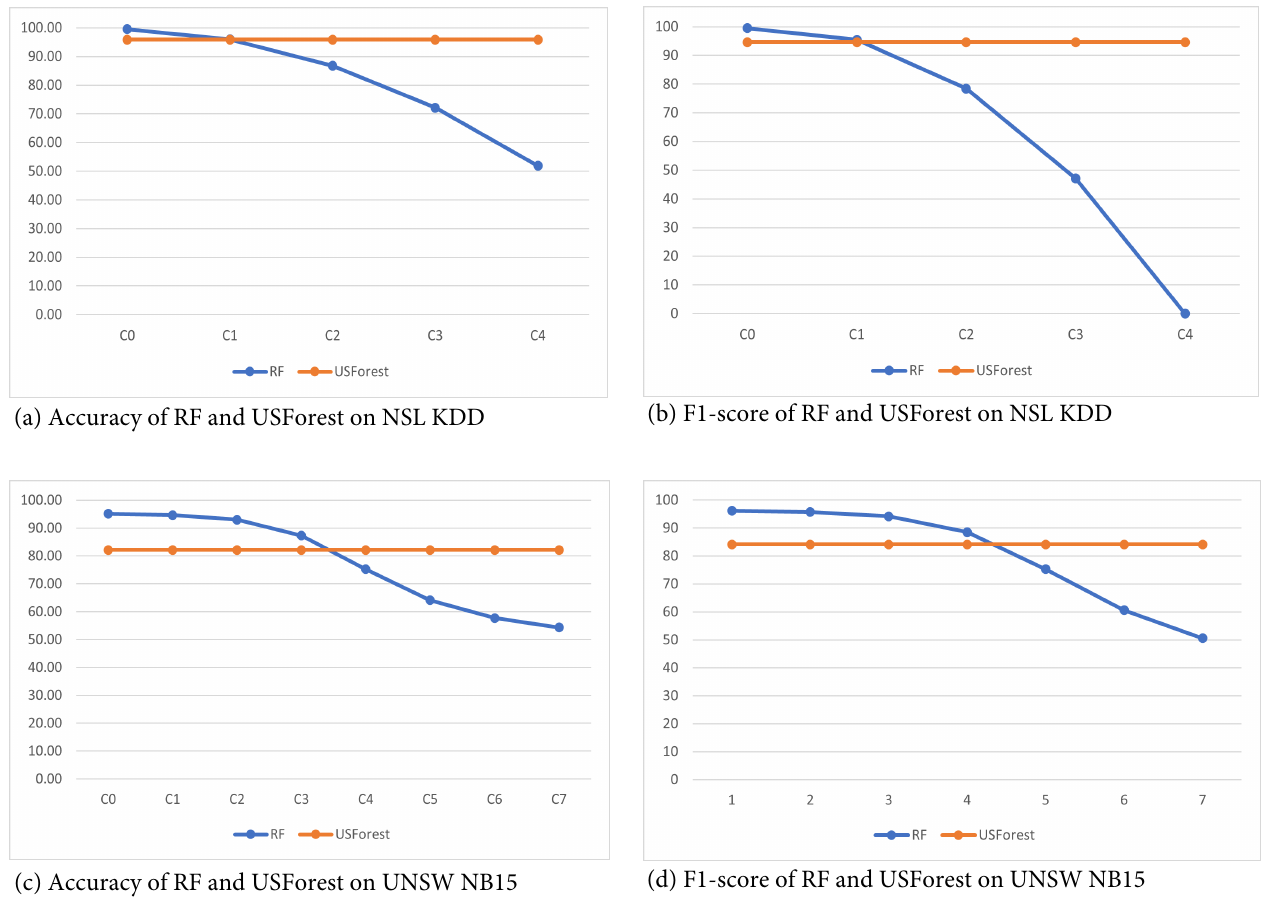}
    \caption{Comparison of RF and usfAD in terms of accuracy and F1-score}
    \label{fig:10}
\end{figure}

\begin{figure}[!htbp]
    \centering
    \includegraphics[scale = .65]{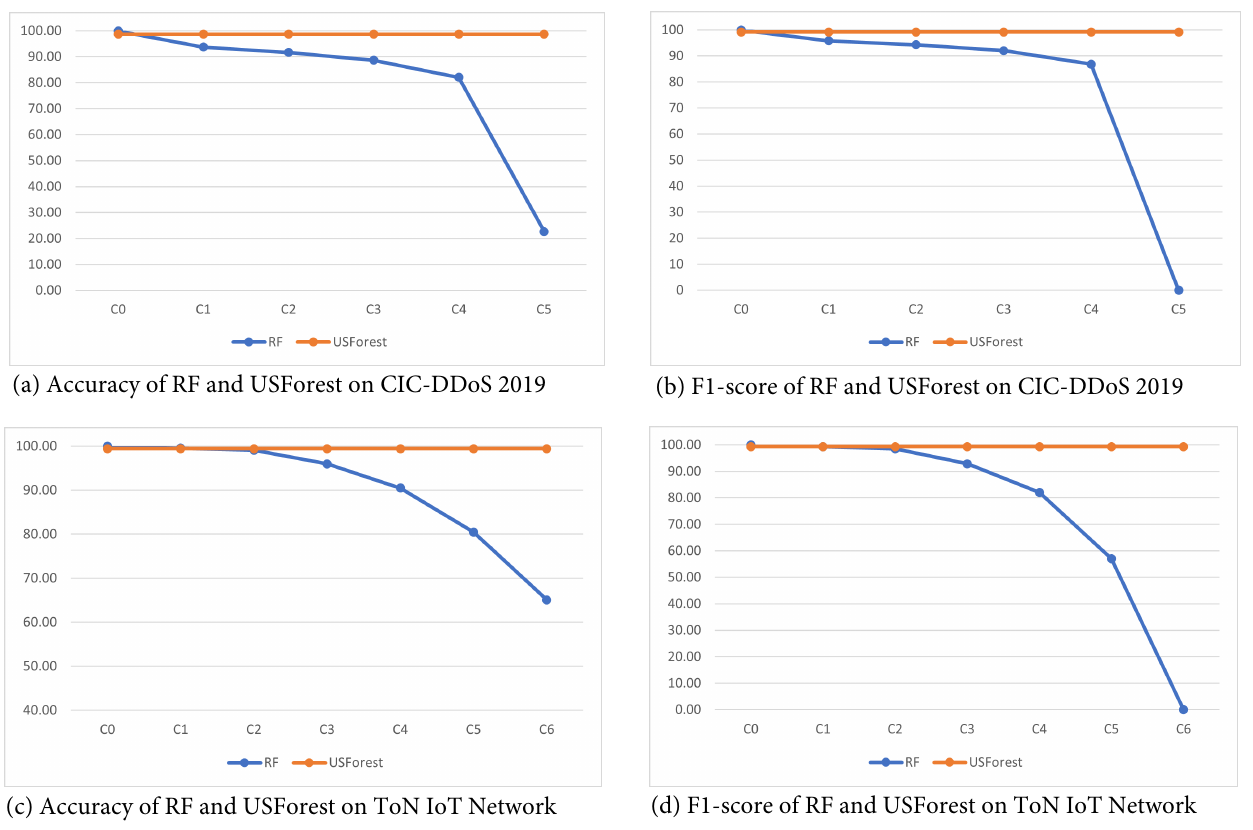}
    \caption{Comparison of RF and usfAD in terms of accuracy and F1-score}
    \label{fig:11}
\end{figure}

Figure \ref{fig:10} and \ref{fig:11} (a) to (d) illustrates the performance of two models: the RF model and the usfAD (an outlier detection method) in terms of their accuracy and F1-score in detecting unknown attack classes for NSL-KDD, UNSW-NB15, CIC-DDoS2019, and ToN-IoT-Network datasets. In graph, C0: The RF model is trained with all types of attack instances that are present in the testing dataset. C1: The RF model is trained excluding one type of attack from the training dataset, which means this particular attack type becomes "unknown" during testing. C2: The RF model is trained excluding two types of attacks from the training dataset, leading to two "unknown" attack types during testing. C3: Similar to the above, the RF model is trained excluding three types of attacks. C4: The RF model is trained excluding four types of attacks and so on. However, in all these scenarios, the usfAD model is trained only on normal instances, meaning it doesn't need knowledge of any attack types during training phase.

\begin{itemize}
    
    \item RF model: In the C0 scenario, the RF model has an extremely high accuracy of 99.49\%, 95.12\%, 99.94\%, 99.99\% (as shown in Figure \ref{fig:7} (a), (c) and \ref{fig:8} (a), (c)) and F1-score of 99.47\%, 96.18\%, 99.92\%, 99.99\% (as shown in Figure \ref{fig:7} (b), (d) and \ref{fig:8} (b), (d)) on NSL-KD, UNSW-NB15, CIC-DDoS2019 and ToN-IoT-Network datasets. This is expected since it is trained with all types of attacks present in the test data. This suggests that its recall and precision are both high when all attack types are known during training. As we move from C1 to C4 for NSL KDD and C1 to C7 for UNSW NB15, we notice a decline in the RF's accuracy and F1-score. This decline corresponds to the increasing number of "unknown" attack types (those that RF wasn't trained on). For NSL KDD, by C4, where RF isn't trained on four types of attacks, its accuracy and F1-score drop to 51.88\% and 0\% respectively.  By C4, the F1-score reduces drastically to 0, implying that the RF model fails completely in terms of both precision and recall for detecting the "unknown" attacks. For other datasets including UNSW NB15, CIC-DDoS2019, and ToN-IoT-Network as we exclude increasing numbers of attack types from the training dataset, we notice a consistent decline in both accuracy and F1-score (C7: 54.36\% and 43.87\% for UNSW NB15, C5: 22.68\% and 0\% for CIC-DDoS2019,  C6: 65.07\%, and 0\% for ToN IoT Network).

    \item usfAD: The performance of usfAD remains consistent across all scenarios with an accuracy of 95.92\%, 82.15\%, 98.69\%, 99.43\% and an F1-score of 94.65\%,84.18\%, 99.19\%, and 99.37\% for NSL KDD, UNSW NB15, CIC-DDoS 2019 and ToN IoT Network. This implies that regardless of the "unknown" attack types in the test data, usfAD can detect anomalies with the same efficiency. usfAD consistently manages to have a balanced precision and recall. This consistent performance can be attributed to the fact that usfAD is trained only on normal instances and focuses on spotting deviations from this norm. 
    
\end{itemize}

While the RF model showed slightly higher accuracy and precision compared to usfAD, it necessitates training with a large number of attack samples and struggles to detect unknown attacks. In contrast, usfAD doesn't need any attack samples, making it better suited for detecting unknown attacks in real-world situations. The results indicate the robustness of the usfAD model in terms of both accuracy and F1-score. Despite the changing landscape of "unknown" attacks in the test data, it manages to consistently perform well. On the other hand, the RF model, while performing admirably when trained with all attack types, sees a dramatic decline in both accuracy and F1-score as more attack types are omitted during training. By the time we reach C4, C7, C5, and C6 for NSL-KDD, UNSW-NB15, CIC-DDoS2019 and ToN-IoT-Network dataset, RF's ability to detect "unknown" attacks deteriorates considerably, emphasizing the challenges of using supervised models when the nature of threats evolves or is not entirely known during training.

\subsection{Comparison of the Proposed Framework with the State-of-the-art Works}

Hairab et al.\cite{hairab2022anomaly} explored three testing scenarios for the BoT-IoT dataset: Scenario A involving normal data and DDoS attacks, Scenario B with normal data and OS Fingerprint attacks, and Scenario C with Normal data and Service Scan attacks. In our research, we included all attack types in the testing set and observed that the recall rate for attacks was 100\%. This finding indicates that there is no need to separate the attack types into distinct scenarios. We achieve the same outcome for all three scenarios. Table \ref{tab:comparison1} shows that the usfAD algorithm (that we first utilized to detect zero-day attacks in IDS) is superior, effectively handling not just specific but all categories of attacks. In addition, our experiments show that the LOF (Local Outlier Factor) also yields comparatively better results, especially in detecting attacks in a zero-day attack scenario than other existing methods on BoT-IoT datasets.

\begin{table}[!htbp]
\centering
\caption{Comparison of our model with Hairab et al.'s model}
\label{tab:comparison1}
\begin{tabular}{|l|l|ll|ll|ll|}
\hline
\multirow{2}{*}{Model} & \multirow{2}{*}{Accuracy} & \multicolumn{2}{l|}{Precision} & \multicolumn{2}{l|}{Recall} & \multicolumn{2}{l|}{F1-score} \\ \cline{3-8} 
 &  & \multicolumn{1}{l|}{Normal} & Attack & \multicolumn{1}{l|}{Normal} & Attack & \multicolumn{1}{l|}{Normal} & Attack \\ \hline
CNN L2-A\cite{hairab2022anomaly} & 99.98 & \multicolumn{1}{l|}{99.96} & 99.99 & \multicolumn{1}{l|}{99.98} & 99.98 & \multicolumn{1}{l|}{99.97} & 99.99 \\ \hline
CNN L2-B\cite{hairab2022anomaly} & 98.49 & \multicolumn{1}{l|}{95.01} & 99.99 & \multicolumn{1}{l|}{99.98} & 97.90 & \multicolumn{1}{l|}{97.43} & 98.93 \\ \hline
CNN L2-C\cite{hairab2022anomaly} & 90.75 & \multicolumn{1}{l|}{75.55} & 99.99 & \multicolumn{1}{l|}{99.98} & 87.05 & \multicolumn{1}{l|}{86.07} & 93.08 \\ \hline
Proposed method-1(LOF) & 98.26 & \multicolumn{1}{l|}{100} & 96.69 & \multicolumn{1}{l|}{96.58} & 100 & \multicolumn{1}{l|}{98.26} & 98.32 \\ \hline
Proposed method-2(usfAD) & 99.66 & \multicolumn{1}{l|}{100} & 99.22 & \multicolumn{1}{l|}{99.21} & 100 & \multicolumn{1}{l|}{99.6} & 99.61 \\ \hline
\end{tabular}
\end{table}

\begin{figure}[!htbp]
    \centering
    \includegraphics[scale = .75]{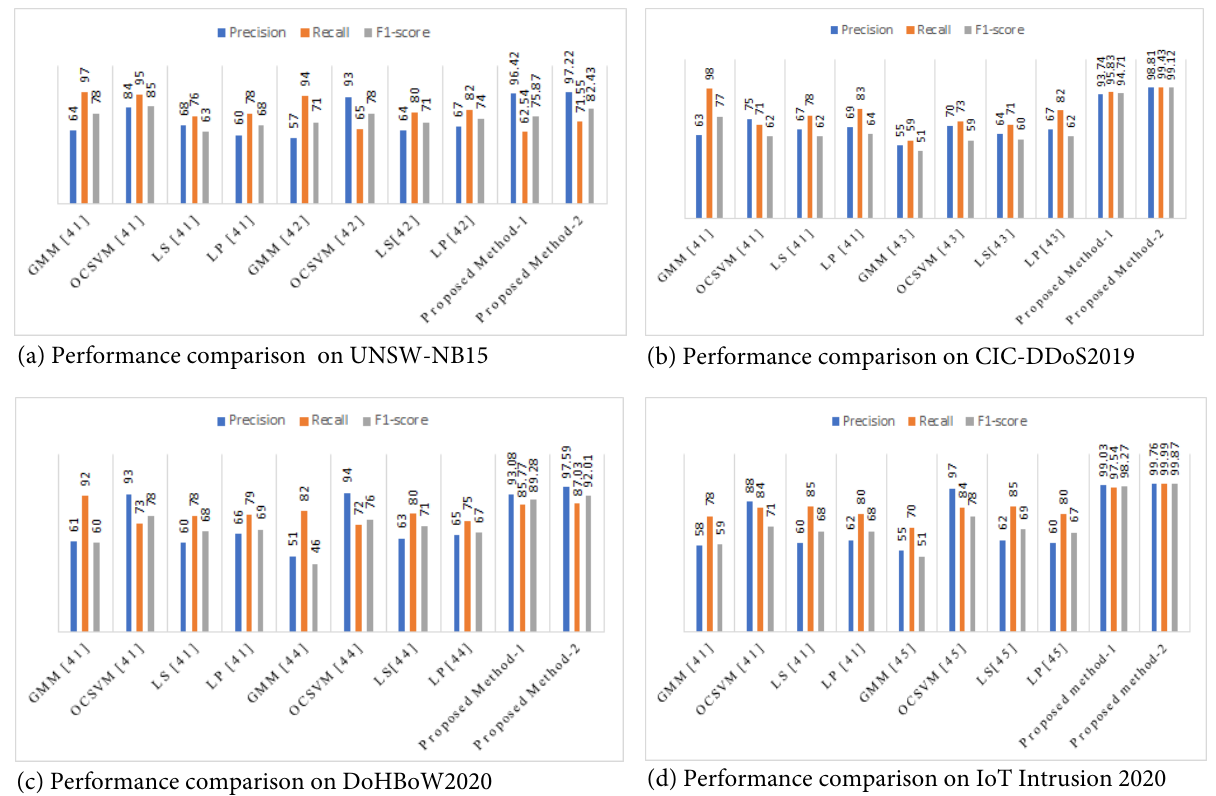}
    \caption{Comparison of the proposed approaches with the state-of-the-art methods}
    \label{fig:12comparison}
\end{figure}

Mbona et al. \cite{mbona2022detecting} applied four methods called Gaussian mixture model, OCSVM, Label spreading and label propagation to detect zero-day attacks in IDS. They utilized four different datasets: UNSW-NB15, CIC-DDoS2019, IoT Intrusion 2020 and CIC-DoHBoW2020 datasets. In Figure \ref{fig:12comparison}, we can observe that except UNSW-NB15 datasets, the proposed method-2(usfAD) outperforms the existing state-of-the-art methods\cite{moustafa2015significant, sharafaldin2019developing, montazerishatoori2020detection}
 in terms of precision, recall and F1-score. Despite state-of-the-art methods showing enhanced performance with the UNSW-NB15 datasets, our experiments revealed relatively lower effectiveness for most techniques, including LOF, OCSVM, and IOF on this datasets.







\begin{figure}[!htbp]
    \centering
    \includegraphics[scale = .85]{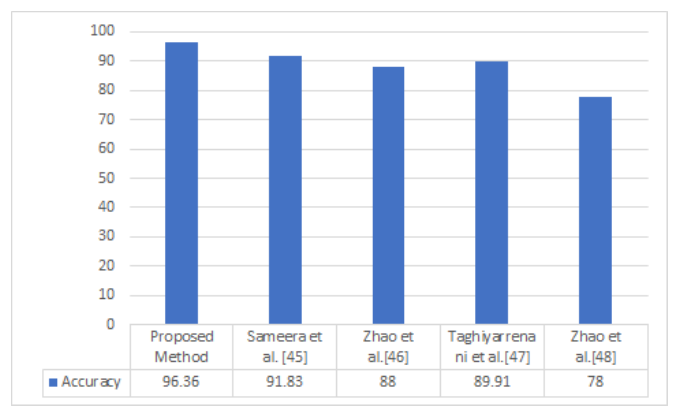}
    \caption{Comparison of the proposed method with the state-of-the-art methods on NSL-KDD}
    \label{fig:13comparison}
\end{figure}

Sameera et al. \cite{sameera2020deep} and other researchers \cite{zhao2019transfer}, \cite{taghiyarrenani2018transfer}, \cite{zhao2017feature} applied deep learning approach, mainly focused on transfer learning to detect zero-day attacks on NSL-KDD datasets. In Figure  \ref{fig:13comparison}, we show the accuracy performance of our method ( accuracy for one of the 10-fold) with the state-of-the-art methods \cite{sameera2020deep}, \cite{zhao2019transfer}, \cite{taghiyarrenani2018transfer}, \cite{zhao2017feature}. We noticed that suggested methods called usfAD outperforms the existing transfer learning approaches on NSL-KDD in terms of accuracy.   


\section{Conclusion}
\label{Conclusion}


In a practical setting, conventional supervised classification based IDS to enforce network security are inappropriate and ineffective because their efficacy depends on the availability of a large number of attack samples. However, it is difficult to acquire attack samples from security-related applications, as the nature of attacks changes frequently. To resolve this issue, we explore two strategies: 1) training supervised learning with datasets of uniformly generated random noise to detect potential future attacks 2) Examine the efficacy of outlier methods and their varied ensemble approaches. Our experiment demonstrated that artificially simulated attack samples with supervised learning are ineffective for detecting unknown attacks. However, our findings indicate that the outlier methods can produce higher accuracy and F1 scores for the majority of benchmark datasets. usfAD is more effective than other widely used outlier techniques for detecting normal and attack classes in an intrusion detection system (IDS). In addition, we demonstrated that a combination of models and ensemble methods could be used to maximize the efficacy of outlier detection in security applications. By selecting and combining models with high recall and precision, it is possible to build a robust system capable of accurately identifying attacks while minimizing false negatives and false positives. Lastly, models such as usfAD and ensemble approach that consistently demonstrate high recall values stand out as effective options for identifying attack instances across a variety of datasets. In our future research, we aim to explore the potential of usfAD for hierarchical multi-classification to identify various cyberattack types, such as DoS, Ransomware, Spyware, and Trojan Horse. In addition, we need to devise an appropriate strategy for generating noise labeled as attacks across the entirety of the feature spaces, which will aid supervised learning in detecting previously unobserved attacks.

\section*{Declarations}

\subsection*{Conflict of interest}
The authors have no conflicts of interest to declare that they are relevant to the content of this article.

\section*{Acknowledgments}
This material is based upon work supported by the Air Force Office of Scientific Research under award number FA2386-23-1-4003.

\section*{Author statements}
Md Ashraf Uddin: Conceptualization; Data curation; Implementation, Roles/Writing-original draft; and Writing, Visualization; Formal analysis.
Sunil Aryal: Funding acquisition; Investigation; Methodology; Project administration; Resources; Software; Supervision; Validation;  Roles/Writing-original draft; and Writing - review \& editing.
Mohamed Reda Bouadjenek: Conceptualization; Project administration;
Muna Al-Hawawreh: Review \& editing.
Md. Alamin Talukder: Data curation; Implementation; Visualization.

\bibliographystyle{els-article}
\bibliography{references}

\end{document}